\documentclass{aastex62}

\graphicspath{{./}{figures/}}

\shorttitle{BSS SEDs in 47~Tuc}
\shortauthors{Raso et al.}

\begin{document}

\title{Spectral Energy Distribution of Blue Stragglers in the core of 47 Tucanae}

\correspondingauthor{Silvia Raso}
\email{silvia.raso2@unibo.it}

\author[0000-0003-4592-1236]{Silvia Raso} \affil{Dipartimento di Fisica e Astronomia, Universit\`a di Bologna, Via Gobetti 93/2, Bologna I-40129, Italy} \affil{Istituto Nazionale di Astrofisica (INAF), Osservatorio di Astrofisica e Scienza dello Spazio di Bologna, Via Gobetti 93/3, Bologna I-40129, Italy}

\author[0000-0002-7104-2107]{Cristina Pallanca} \affil{Dipartimento di Fisica e Astronomia, Universit\`a di Bologna, Via Gobetti 93/2, Bologna I-40129, Italy}  \affil{Istituto Nazionale di Astrofisica (INAF), Osservatorio di Astrofisica e Scienza dello Spazio di Bologna, Via Gobetti 93/3, Bologna I-40129, Italy}

\author[0000-0002-2165-8528]{Francesco R. Ferraro} \affil{Dipartimento di Fisica e Astronomia, Universit\`a di Bologna, Via Gobetti 93/2, Bologna I-40129, Italy}  \affil{Istituto Nazionale di Astrofisica (INAF), Osservatorio di Astrofisica e Scienza dello Spazio di Bologna, Via Gobetti 93/3, Bologna I-40129, Italy}

\author[0000-0001-5613-4938]{Barbara Lanzoni} \affil{Dipartimento di Fisica e Astronomia, Universit\`a di Bologna, Via Gobetti 93/2, Bologna I-40129, Italy}  \affil{Istituto Nazionale di Astrofisica (INAF), Osservatorio di Astrofisica e Scienza dello Spazio di Bologna, Via Gobetti 93/3, Bologna I-40129, Italy}

\author[0000-0001-9158-8580]{Alessio Mucciarelli} \affil{Dipartimento di Fisica e Astronomia, Universit\`a di Bologna, Via Gobetti 93/2, Bologna I-40129, Italy}  \affil{Istituto Nazionale di Astrofisica (INAF), Osservatorio di Astrofisica e Scienza dello Spazio di Bologna, Via Gobetti 93/3, Bologna I-40129, Italy}

\author[0000-0002-6040-5849]{Livia Origlia}\affil{Istituto Nazionale di Astrofisica (INAF), Osservatorio di Astrofisica e Scienza dello Spazio di Bologna, Via Gobetti 93/3, Bologna I-40129, Italy}

\author[0000-0003-4237-4601]{Emanuele Dalessandro} \affil{Istituto Nazionale di Astrofisica (INAF), Osservatorio di Astrofisica e Scienza dello Spazio di Bologna, Via Gobetti 93/3, Bologna I-40129, Italy}

\author[0000-0003-3858-637X]{Andrea Bellini} \affil{Space Telescope Science Institute, 3700 San Martin Drive, Baltimore, MD 21218, USA}

\author[0000-0001-9673-7397]{Mattia Libralato} \affil{Space Telescope Science Institute, 3700 San Martin Drive, Baltimore, MD 21218, USA}

\author[0000-0003-2861-3995]{Jay Anderson} \affil{Space Telescope Science Institute, 3700 San Martin Drive, Baltimore, MD 21218, USA}

\begin{abstract}

We have constructed the Spectral Energy Distributions (SEDs) of a sample of Blue Straggler Stars (BSSs) in the core of the globular cluster 47~Tucanae, taking advantage of the large set of high resolution images, ranging from the ultraviolet to the near infrared, obtained with the ACS/HRC camera of the \textit{Hubble Space Telescope}. Our final BSS sample consists of 22 objects, spanning the whole color and magnitude extension of the BSS sequence in 47~Tucanae. We fitted the BSS broadband SEDs with models to derive temperature, surface gravity, radius, luminosity and mass. We show that BSSs indeed define a mass sequence, where the mass increases for increasing luminosity. Interestingly, the BSS masses estimates from the SED fitting turn out to be comparable to those derived from the projection of the stellar position in the color-magnitude diagram onto standard star evolutionary tracks. We compare our results with previous, direct mass estimates of a few BSSs in 47~Tucanae. We also find a couple of supermassive BSS candidates, i.e., BSSs with a mass larger than twice the turn-off mass, the formation of which must have involved more than two progenitors.

\end{abstract}


\section{Introduction} \label{sec:1}
In the color-magnitude diagram (CMD) of a stellar population, Blue Straggler Stars (BSSs) are a peculiar group of stars that lies along an extrapolation of the main sequence (MS), at brighter magnitudes and bluer colors with respect to the turn-off (TO) point (e.g., \citealt{sandage53, ferraro92, ferraro93, ferraro97, ferraro04, ferraro18, piotto04, lanzoni07a, lanzoni07b, leigh07, dalessandro08, moretti08, beccari11, beccari12, simunovicpuzia16}).
They are the heaviest luminous (non-degenerate) stars in a GC, as suggested by their position in the CMD and as confirmed by a few mass measurements (e.g., \citealt{shara97, gilliland98, demarco05, ferraro06, fiorentino14, baldwin16, libralato18, libralato19}). They essentially mimic a younger population, although in GCs any recent star formation event can be safely excluded. Therefore, to explain their presence we must take into account mechanisms able to increase stellar mass. Two main formation processes have been proposed so far: mass-transfer in binary systems (\citealt{mccrea64}) and direct collisions between stars (\citealt{hillsday76}). 
Although few pieces of evidences suggest that both mechanisms can be active simultaneously in the same GC (\citealt{ferraro09, dalessandro13, simunovic14, xin15, beccari19}), a complete understanding of their relative efficiency and their dependence on the parent cluster properties is still lacking (e.g., \citealt{ferraro95, ferraro03, davies04, sollima08, chenhan09, knigge09, chatterjee13, leigh13, sills13}).

However, irrespective of their origins, BSSs have been demonstrated to be very powerful indicators of GC internal dynamics (\citealt{ferraro12,ferraro18,lanzoni16}). This is because they are significantly more massive ($M_{\rm BSS}\sim 1.2$-$1.3 M_\odot$) than the average stellar mass in GCs ($\langle m\rangle \sim 0.4 M_\odot$), and they are therefore subject to dynamical friction, which makes them sink toward the cluster center (e.g., \citealt{mapelli04,mapelli06}).
Since dynamical friction primarily depends on the local density (other than on the BSS mass; e.g., \citealt{alessandrini14}), it affects BSSs orbiting at progressively larger distances from the center, continuously modifying their radial distribution within the host cluster. The comparison between the shape of the BSS radial distribution and that of normal (lighter) cluster stars has thus been used to define the so-called ``dynamical clock'' (\citealt{ferraro12}) and, more recently, the $A^+$ parameter (\citealt{alessandrini16}), that allow an empirical ranking of stellar systems based on their dynamical age, i.e., the stage of internal dynamical evolution they achieved (\citealt{lanzoni16,raso17,ferraro18,dalessandro18,li19,singhyadav19,sollimaferraro19}). Indeed, by investigating $\sim 30\%$ of the entire GC population in the Milky Way, a strong correlation has been found between the value of $A^+$ and the number of current central relaxation times occurred since the epoch of cluster formation (\citealt{ferraro18}), thus solidly demonstrating the effectiveness of this parameter as dynamical indicator. In this respect an accurate knowledge of the BSS mass distribution would be of paramount importance for a precise estimate of the BSS sedimentation timescale, thus allowing an accurate calibration of these empirical tools. Moreover, this would also provide crucial hints for a deeper understanding of the formation mechanisms and evolutionary processes of these puzzling stars.
In spite of their importance, BSS masses have been determined only for a few BSSs per clusters (e.g., \citealt{shara97, gilliland98, demarco05, fiorentino14}). These sparse measurements have helped to confirm  that BSSs are indeed more massive than MS stars. However, systematic studies aimed at obtaining direct mass estimates for large samples of BSSs, covering the entire extension of the sequence, are still unavailable due to observational difficulties.

BSS mass measurements can be obtained spectroscopically (e.g., \citealt{shara97, demarco05}; see also the recent estimate of the mass of an evolved BSS in 47 Tuc by \citealt{ferraro16}), from pulsational properties (e.g., \citealt{gilliland98, fiorentino14}), or through spectral energy distribution (SED) fitting (e.g., \citealt{knigge06, knigge08}; in the latter case, also combined with far-ultraviolet, hereafter FUV, spectroscopy). All these methods require specifically designed observations and techniques: spectroscopic observations must deal with serious crowding issues in the dense environment of GCs, especially in their cores; variability and SED-based studies require \textit{ad hoc} datasets (time-series photometry for the former, and photometry in a large number of filters for the latter), both of which are rarely available in the archives.

The GC 47~Tucanae (47~Tuc, NGC~104) is a notable exception, since it has been intensively studied over the years and has also been used as a calibration field for different \textit{Hubble Space Telescope} (\textit{HST}) instruments. Therefore, a large and multi-band set of images of the core of 47 Tuc is publicly available (see Sect.~\ref{s:2} for a detailed description of the dataset). In this work, we use the Advanced Camera for Surveys (ACS) High Resolution Channel (HRC) images in this dataset to construct broad (extending from $2000$ to $8000$ \AA) SEDs for a sample of 22 BSSs, distributed along the entire extent of the BSS sequence in 47~Tuc. We used these SEDs to derive BSS physical properties (such as luminosities, masses, radii, etc.).

The paper is organized as follows: in Sect.~\ref{s:2} we describe the dataset and the data reduction procedure; in Sect.~\ref{s:3} we select the BSS sample and describe the SED-fitting procedure we used; in Sect.~\ref{s:4} we discuss our results, and in Sect.~\ref{s:5} we summarize our conclusions.

\section{Dataset and data reduction}\label{s:2}

\begin{figure*}[!t]
\centering
\includegraphics[width=0.95\textwidth]{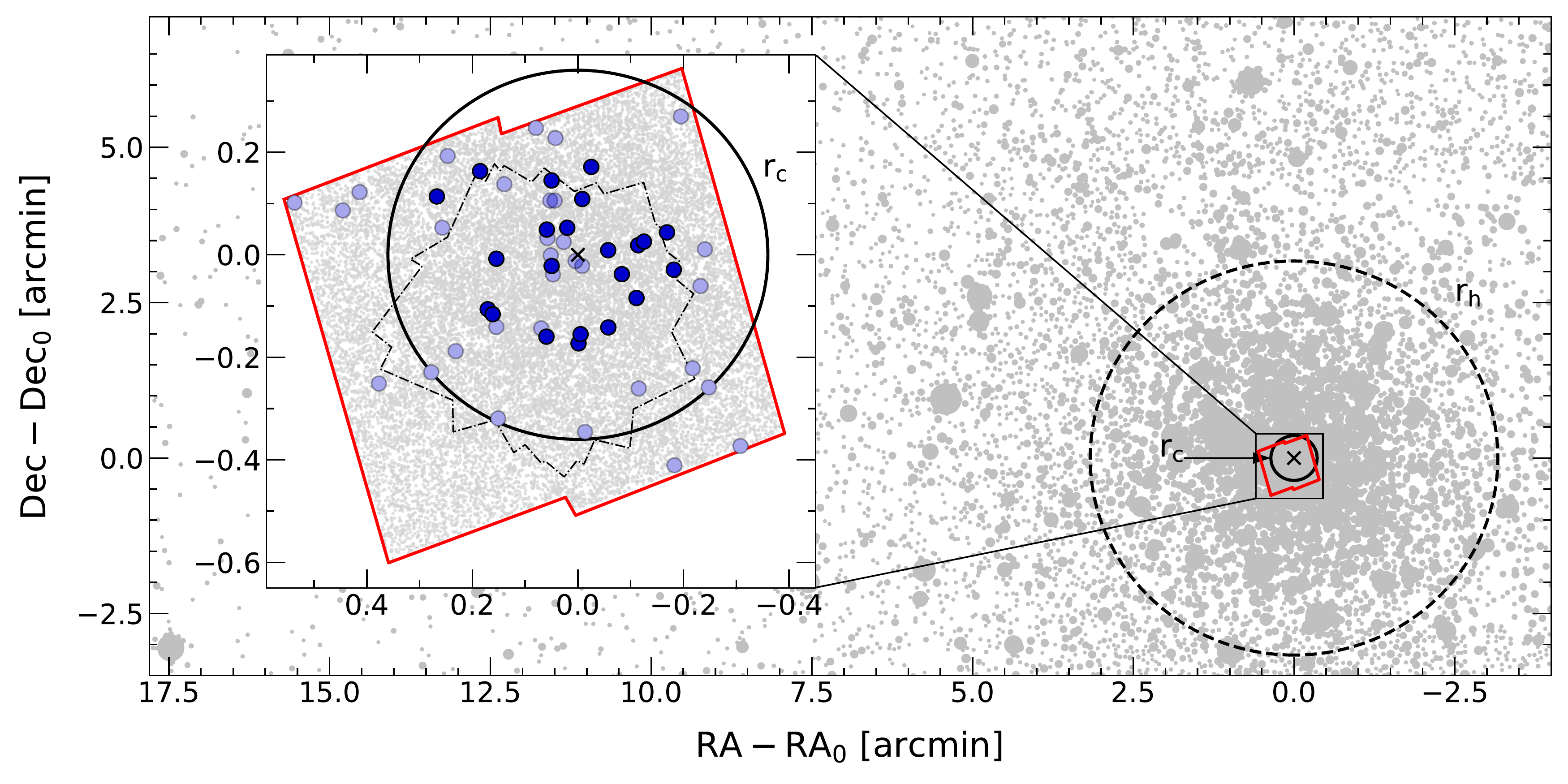}
\caption{
Map of the 47~Tuc region analyzed in this work. Positions are in arcmin, with respect to the cluster center (small black cross). The solid and dashed circles represent the core and half-light radius of the cluster, respectively, from \citet{h96}. The red contour marks the FOV of the ACS/HRC dataset used in this work. In the inset on the left, we zoom in the ACS/HRC FOV. The dotted-dashed contour marks the FOV where the time baseline is long enough to measure PMs. The dark blue circles represent the BSSs that survived our selections (as described in Sect.~\ref{ss:samplesel}), while the light-blue circles are all the remaining BSSs detected within the FOV.}
\label{fig:FOV}
\end{figure*}

\begin{table*}[t!]
\caption{List of \textit{HST} ACS/HRC observations of 47~Tuc used in this work.\label{tab:1}}
\centering
{
\begin{tabular}{lllll}
\hline\hline
Program ID & PI & Epoch & Filter & Exposures \\
 & & (yyyy/mm) & & N $\times~t_{exp}$ \\
\hline
\hline
9019 & R. Bohlin & 2002/04     & F220W & $21\times 170\,{\rm s}$\\
	 &		   &		       & F330W & $18\times 66\,{\rm s}$ \\
	 &		   &		       & F435W & $2\times 5\,{\rm s}, 2\times 20\,{\rm s}, 17\times 60\,{\rm s}, 2\times 300\,{\rm s}  $ \\
	 &		   &		       & F475W & $10\times 60\,{\rm s}$ \\
	 &		   &		       & F555W & $14\times 60\,{\rm s}  $ \\
	 &		   &		       & F606W & $10\times 60\,{\rm s}$ \\
	 &		   &		       & F625W & $10\times 60\,{\rm s}$ \\
	 &		   &		       & F775W & $13\times 60\,{\rm s}$ \\
	 &		   &		       & F814W & $2\times 5\,{\rm s}, 2\times 20\,{\rm s}, 14\times 60\,{\rm s} $\\
\hline
9028 & G. Meurer & 2002/04     & F475W & $40\times 60\,{\rm s}$\\
\hline
9443 & I. King & 2002/07    & F250W & $1\times 230\,{\rm s}, 1\times 460\,{\rm s}$\\
	 &		   &		  & F330W & $1\times 350\,{\rm s}$ \\
	 &		   &		  & F435W & $1\times 350\,{\rm s}$ \\
	 &		   &		  & F475W & $20\times 60\,{\rm s}$ \\
\hline
9662 & R. Gilliland & 2002/09  & F606W & $2\times 1\,{\rm s}$\\
\hline
10055 & J. Biretta & 2004/02  & F250W & $2\times 50\,{\rm s}$\\
	  &		    &		  		& F330W & $2\times 40\,{\rm s}$ \\
	  &		    &		  		& F435W & $2\times 20\,{\rm s}$ \\
	  &		    &		  		& F606W & $2\times 10\,{\rm s}$ \\
	  &		    &		  		& F775W & $2\times 10\,{\rm s}$ \\
\hline
10375 & J. Mack & 2004/12; 2005/03; 2005/06; 2005/10  & F435W & $4\times 60\,{\rm s}$\\
	  &		   &		  							   & F475W & $4\times 60\,{\rm s}$\\
	  &		   &		  							   & F555W & $4\times 60\,{\rm s}$\\
	  &		   &		  							   & F606W & $4\times 60\,{\rm s}$\\
	  &		   &		  							   & F625W & $4\times 60\,{\rm s}$\\
	  &		   &		  							   & F775W & $4\times 60\,{\rm s}$\\
	  &		   &		  							   & F814W & $4\times 60\,{\rm s}$\\

\hline
10737 & J. Mack & 2006/03; 2006/05; 2006/07  & F330W & $2\times 66\,{\rm s}$\\
	  &		   &                            & F435W & $6\times 60\,{\rm s}$ \\
	  &		   &		  		& F475W & $6\times 60\,{\rm s}$ \\
	  &		   &		  		& F555W & $6\times 60\,{\rm s}$ \\
	  &		   &		  		& F606W & $6\times 60\,{\rm s}$ \\
	  &		   & 		  		& F625W & $6\times 60\,{\rm s}$ \\
	  &		   &		  		& F775W & $6\times 60\,{\rm s}$ \\
	  &		   &		  		& F814W & $6\times 60\,{\rm s}$ \\

\hline
\hline
\end{tabular}}
\end{table*}

The innermost region of 47 Tuc was used as calibration field for the ACS/HRC, e.g., to study flat-field stability and geometric distortion. For this reason, a large photometric dataset is available in the \textit{HST} archive for this region of the sky, which has been repeatedly observed throughout the years of operation of the ACS/HRC (2002-2007). From the archive, we selected all images obtained through broadband filters, which range from the near UV to the near infrared (approximately, from 2000~\AA~to~8000~\AA), thus covering the whole spectral extension of a typical BSS. The total field of view (FOV) covered by this dataset is shown in Figure~\ref{fig:FOV} (red contour), compared to the core and half-light radii of the cluster (0.36 and 3.17 arcmin, respectively; \citealt{h96}). In Table~\ref{tab:1} we list the images used for this study; our final dataset consists of 285 images. 
Nearly the same data were used by \citet{knigge06, knigge08} to study exotic objects such as cataclysmic variables (CVs), white dwarfs (WDs) and BSSs. In particular, physical parameters of three BSSs were obtained from spectra and SED fitting. We
compare their results with ours in Sect.~\ref{ss:compmass}.

The photometric reduction was performed on the \texttt{\_flt} exposures, because the un-resampled pixel data for stellar profile fitting is preserved. We followed the procedures described in \citet{2017ApJ...842....6B, bellini18}, which we briefly summarize in the following.

First, we performed one-pass photometry on the images. This consisted of a single finding procedure without neighbour subtraction. 
For this step, we determined a spatially variable PSF model from each individual exposure by examining the residuals from the fit of an empirical library PSF (see \citealt{ak04}) to the bright, relatively isolated, unsaturated stars in that exposure. We then used this tailor-made PSF to measure stellar positions and fluxes in each exposure.
We corrected stellar positions for geometric distortion using the distortion solutions provided by \citet{ak04}.
 
Since we are focusing on the central, most crowded regions of the cluster, we secondly used a multi-pass photometric procedure, which is able to perform neighbour subtraction. We used the stellar positions in the early release catalog from the \textit{HST UV Legacy Survey of Galactic Globular Clusters} (\citealt{piotto15, soto17}) as an absolute astrometric reference system. Based on these RA and Dec positions, we defined a common, pixel-based reference system, with the X and Y axes increasing, respectively, toward West and North, and with the center of the cluster arbitrarily placed at position (5000,5000). We set the pixel scale to be 25 mas pixel$^{-1}$, consistent with that of ACS/HRC. We transformed each stellar position from the single-exposure, one-pass photometry catalogs into the reference frame by means of six-parameter linear transformations, using a subset of bright, unsaturated and well-measured stars in common between the two catalogs. We rescaled the instrumental magnitudes of each exposure to match those of the longest available exposures taken with that filter. 

The multi-pass photometry was performed with the code \texttt{KS2} (see \citealt{2017ApJ...842....6B} for details). \texttt{KS2} combines the results of the one-pass photometry, transformed into the common reference frame, and it is able to simultaneously analyze multiple images of a given stellar field obtained with different filters.
For this study, differently from the UV-driven approach adopted in \citet{raso17}, we performed the finding procedure simultaneously on all the available exposures. 
This different approach is motivated by the necessity of detecting and subtracting all the potential contaminators (even faint red stars) that can potentially affect the SED of each selected BSS. Note that in the case of the data analyzed here, crowding is reduced because of the very high angular resolution of the ACS/HRC and the relative proximity of the cluster ($D=4.5$ kpc, \citealt{h96}).

\texttt{KS2} also provides a set of photometric quality parameters that can be used to select well-measured stars (see \citealt{2017ApJ...842....6B} for a complete description). 
Briefly, these parameters are: the \texttt{QFIT} parameter, which indicates the quality of the PSF fit; the \textit{o} parameter, which measures the neighbour flux (normalized to the star flux) that had to be subtracted within the fitting radius; and the \texttt{RADXS} parameter, which measures the source flux beyond the fitting radius, with respect to the flux predicted by the PSF. The \texttt{RADXS} parameter is useful to distinguish between extended sources, like galaxies or blends, which have a substantial excess of flux outside the fitting radius with respect to the PSF, and cosmic rays or hot pixels, which have less flux in the outskirts of their profile with respect to the PSF predictions.

Our dataset covers a time baseline of about 4 years, long enough to measure proper motions (PMs), in order to separate cluster stars from field stars. We measured PMs using the technique developed in \citet{b14} and recently improved in \citet{bellini18}. The region where PMs can be measured is smaller than the FOV of the whole dataset (see the dashed-dotted contour in the inset of Figure~\ref{fig:FOV}), since the external regions were only observed in one epoch (as part of the observing programs 9019 and 9028 performed in 2002).

\subsection{Photometric calibration}\label{ss:2.1}

\begin{table}[t!]
\caption{List of photometric calibration values.\label{tab:2}}
\centering
{
\begin{tabular}{llll}
\hline\hline
Filter& $\Delta \mathrm{mag}$& $\epsilon_{\Delta \mathrm{mag}}$ & $\mathrm{ZP}_{f}$\\
\hline
\hline
F220W & 5.336 & 0.029 & 21.883 \\
F250W & 5.782 & 0.042 & 22.261 \\
F330W & 4.407 & 0.025 & 22.913 \\
F435W & 4.342 & 0.030 & 25.188 \\
F475W & 4.398 & 0.074 & 25.635 \\
F555W & 4.391 & 0.054 & 25.261 \\
F606W & 4.433 & 0.056 & 25.906 \\
F625W & 4.394 & 0.062 & 25.210 \\
F775W & 4.305 & 0.051 & 24.568 \\
F814W & 4.297 & 0.078 & 24.856 \\
\hline

\hline
\end{tabular}}
\end{table}

Since the main goal of this work is to obtain broadband SEDs for a sample of BSSs and use them to estimate their physical parameters, a careful photometric calibration is required.
Following the prescriptions given in \citet{2017ApJ...842....6B}, we obtained \texttt{VEGAMAG} calibrated magnitudes from our instrumental magnitudes as follows:

\begin{equation}\label{eq:1}
    m_{f,\mathrm{CAL}} = m_{f, \mathrm{INSTR}} + \Delta \mathrm{mag} + \mathrm{ZP}_{f}
\end{equation}
where $m_{f,\mathrm{CAL}}$ is the calibrated magnitude in the \texttt{VEGAMAG} system in the considered filter \textit{f}; $m_{f, \mathrm{INSTR}}$ is the instrumental magnitude resulting from the multi-pass photometry; $\Delta \mathrm{mag}$ is the $2.5\sigma$-clipped median difference between the aperture photometry $m_{\mathrm{AP}} (\lambda)$ and the instrumental magnitudes; $\mathrm{ZP}_{f}$ is the photometric zero point of the filter considered\footnote{The photometric zero-points were obtained using the ACS zeropoints calculator available at \url{https://acszeropoints.stsci.edu/}}. 
The value $m_{\mathrm{AP}} (\lambda)$ is measured on the \texttt{\_drz} images using aperture photometry with a 6-pixel radius and corrected for the finite aperture using the encircled energy values listed in \citet{bohlin16}.
We chose to use a 6-pixel aperture since it represents the best compromise between the need to minimize the contribution of nearby stars and the need for a large aperture.
In Table~\ref{tab:2} we list the values of $\Delta \mathrm{mag}$, with their errors $\epsilon_{\Delta \mathrm{mag}}$, and $\mathrm{ZP}_{f}$ used in this work. 

In order to compute the global photometric error for each star, we combined (in quadrature) the rms of the stellar mean magnitude with the uncertainties of the calibration process. 
The dominant component in the calibration error budget comes from the $\Delta \mathrm{mag}$ rms, which is of the order of $10^{-2}-10^{-1}$ mag, while we neglected a minor possible contribution (of the order of $10^{-3}$ mag) due to the variation of the $\mathrm{ZP}_{f}$ as a function of time.

\section{BSS selection and SED-fitting procedure}\label{s:3}

\subsection{BSS Selection}\label{ss:samplesel}

\begin{figure}[!t]
\centering
\includegraphics[width=.60\textwidth]{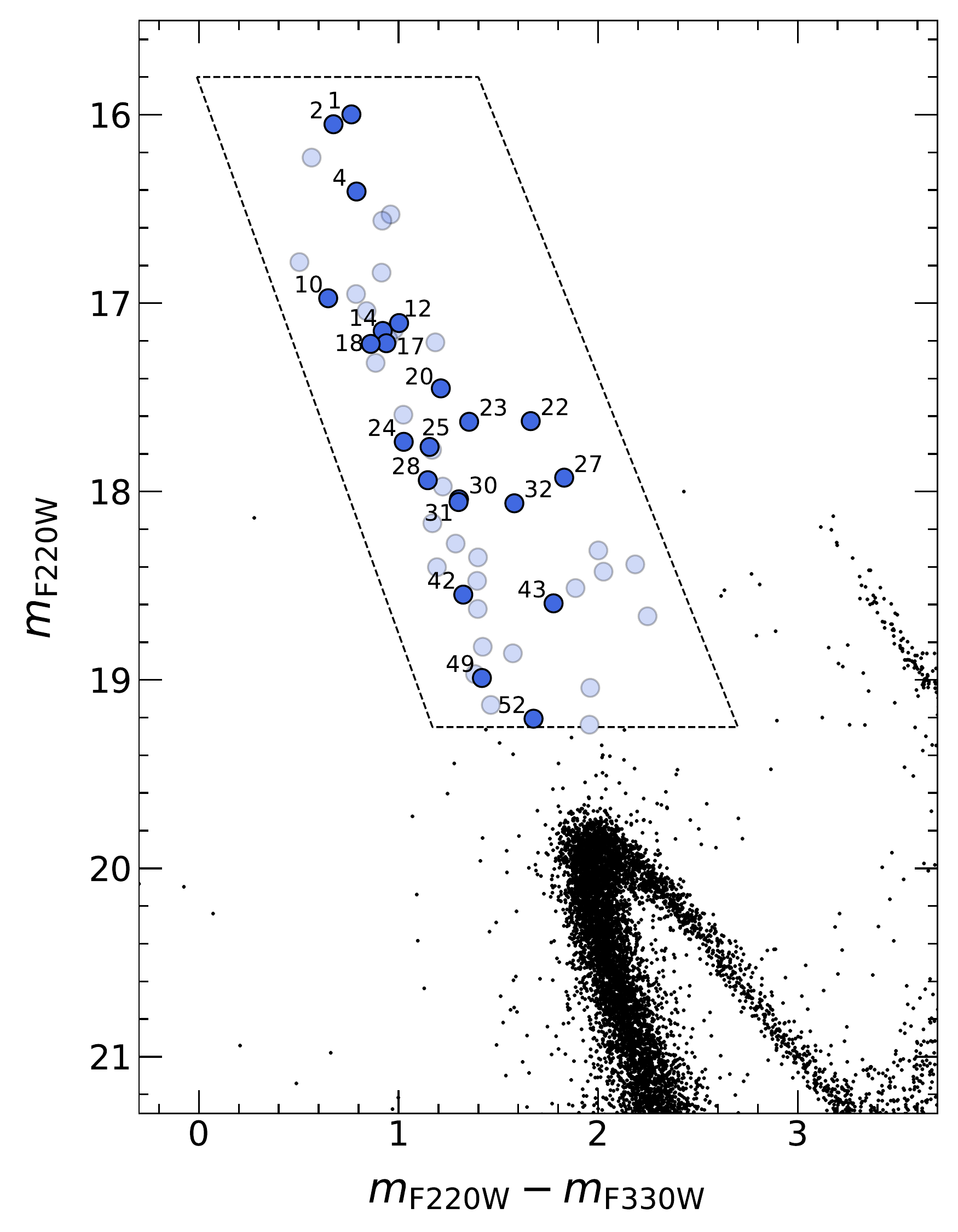}
\caption{UV CMD ($m_{\mathrm{F220W}}$ vs. $m_{\mathrm{F220W}}-m_{\mathrm{F330W}}$) of the central region of 47~Tuc. The dashed box define the region where we selected BSSs. The blue circles (both light and dark) are the complete sample of BSSs in our dataset (53 objects), while the dark blue circles represent the cleaned BSS sample (22 objects, see text for details on the selection). The dark blue circles are labeled as in Table~\ref{tab:results}.}\label{fig:CMDsel}
\end{figure}

Various studies have shown that BSSs in 47~Tuc are concentrated towards the cluster center, as expected for a population of stars heavier than the average (e.g., \citealt{paresce91, Guhathakurta92, demarchi93, ferraro01, ferraro04, ferraro12, parada16b}). Hence, a significant sample of BSSs is expected to fall in the studied FOV, which probes the innermost $50\times50$ arcsec$^2$ of the cluster (see Figure~\ref{fig:FOV}).

We used an ultraviolet CMD ($m_{\mathrm{F220W}}$ vs. $m_{\mathrm{F220W}}-m_{\mathrm{F330W}}$) to select BSSs (similar to the approach used in \citealt{ferraro01, raso17}). In these UV filters, BSSs are among the brightest objects in the cluster, and they define a clear, almost vertical sequence, easily distinguishable from other stellar populations. We selected as BSSs all the 53 stars that fall within the dashed box reported in Figure~\ref{fig:CMDsel}.

Only a sub-sample of the selected BSSs, represented as dark blue circles in Figure~\ref{fig:CMDsel}, has been used to study the SEDs. First of all, we rejected stars that were detected in less than 8 out of the 10 available bandpasses, in order to have a significant number of spectral points for the SED-fitting procedure (see Sect.~\ref{ss:sedfitting}). We also excluded from our sample stars that were measured in less than 2 single exposures per filter, in an effort to include in the final sample only BSSs with robustly measured magnitudes. A further selection was performed by using the quality parameters obtained from the reduction software (described in Sect.~\ref{s:2}).
For each BSS, we computed the median value, over all the filters available, of the photometric error, \texttt{QFIT} and \texttt{RADXS} parameters. These median values do not have a real physical meaning, since they are averaged over different filters, but still they provide an overall photometric quality assessment. For example, an extended source should have a large \texttt{RADXS} value regardless of the filter.
We thus computed the median and the relative error of the three parameters for all the 53 BSSs in our sample, and we assumed these values in order to select well-measured BSSs by rejecting any star having at least one parameter exceeding $5\sigma$ the mean value. We arbitrarily assumed $\textit{o}=0.2$ (i.e., the median neighbour flux subtracted before measuring the star was equal to the $20\%$ of the star flux itself) as our fixed rejection threshold, to safely exclude stars with bright neighbours.

\begin{figure*}[!t]
\centering
\includegraphics[width=.55\textwidth]{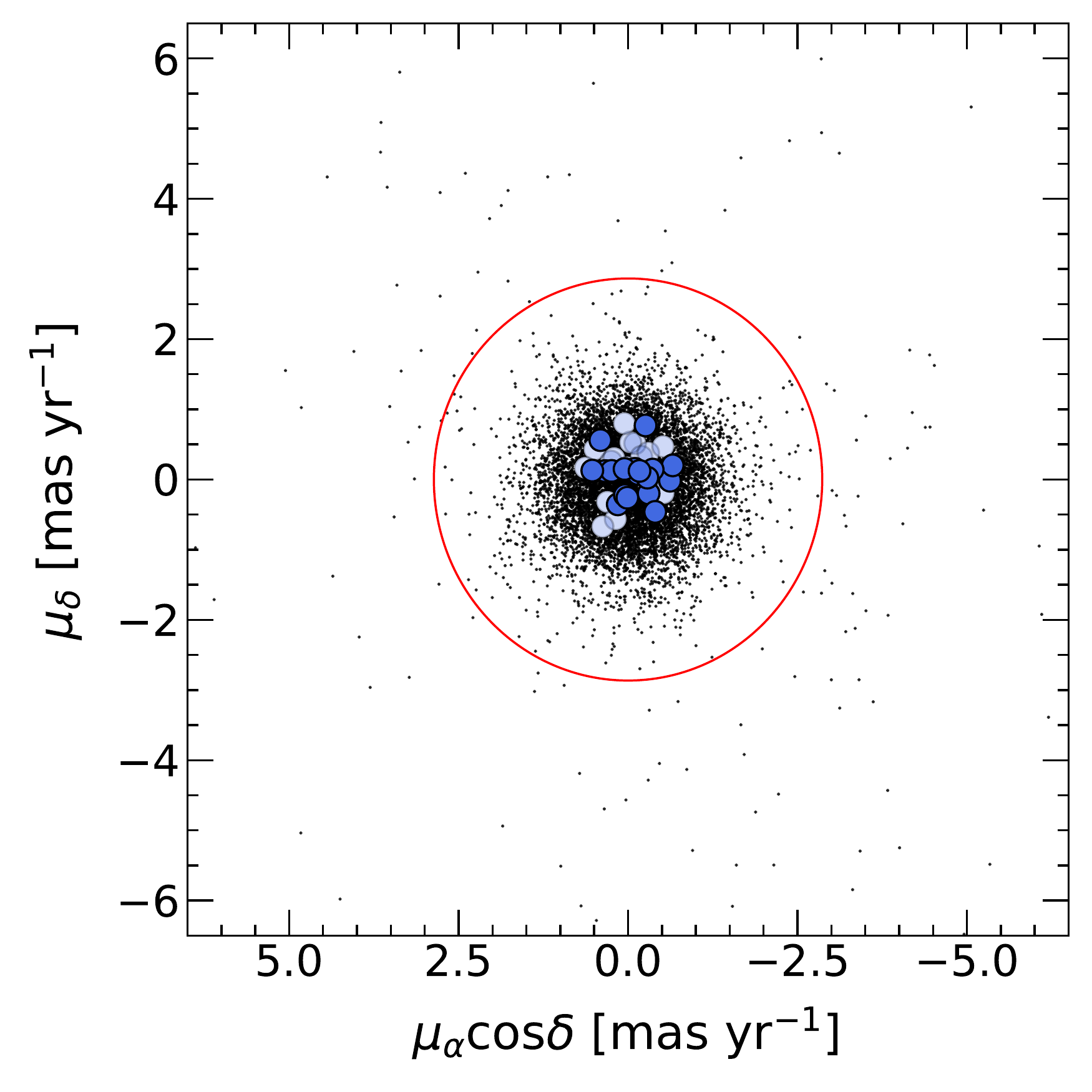}
\caption{Vector-point diagram for all the stars with a PM measurement in our dataset (black points). The blue circles (color coded as in Figure~\ref{fig:CMDsel}) correspond to the 34 BSSs for which PMs could be measured. The red circle corresponds to 5 times the central velocity dispersion of bright stars in 47~Tuc, plotted as a reference.}
\label{fig:vpd}
\end{figure*}

We also checked the cluster membership of the selected BSSs. We show in Figure~\ref{fig:vpd} the vector-point diagram (VPD) for all the stars with a valid PM measurement (black points). As expected, the figure shows a well-defined clump of stars (indicating the cluster members) with essentially no evidence of field contamination: this was somehow expected,  since the observations sampled the innermost regions of the cluster. For reference, the red circle in Figure~\ref{fig:vpd} corresponds to 5 times the central velocity dispersion of bright stars in 47~Tuc (0.573 mas yr$^{-1}$, or 12.2 km s$^{-1}$; \citealt{watkins15, baumgardthilker18}). 
Since the FOV where PMs can be measured is smaller than the FOV of the whole dataset (see Figure~\ref{fig:FOV}), we can provide PM information for only 34 (out of 53) BSSs in our sample (both the light and dark blue circles in Figure~\ref{fig:vpd}; the dark blue circles represent BSSs that also survived the quality and variability selection). 
All these 34 BSSs are cluster members, as can be seen from Figure~\ref{fig:vpd}. Given the negligible level of field contamination, also the remaining 19 BSSs, for which we do not have a PM measurement, have a high probability of being cluster members.

\subsection{Variability}\label{ss:variab}

Variable stars should be excluded from the final BSS sample to avoid the construction of SEDs using magnitudes measured at different phases of variability. The BSS sequence crosses the classical instability strip, so we expect that a few BSSs in our sample could be pulsating variables (e.g., SX Phoenicis). Moreover, some of them are known or suspected eclipsing variables or contact binaries (e.g., W Uma variables). Hence, to perform a meaningful comparison with theoretical SEDs, we excluded all variable BSSs by cross-correlating the BSS positions in our catalog with the positions of variable objects listed in the catalog of Variable Stars in Galactic Globular Clusters (\citealt{clement01varcat}, last update for 47~Tuc: January 2017\footnote{The catalog is available at \url{http://www.astro.utoronto.ca/~cclement/cat/C0021m723}.}), to identify the already known or suspected variables (both eclipsing and pulsating). We focused our attention on variables from \citet{edmonds96} and \citet{gilliland98}, who studied variability in the central $66\times66$ arcsec$^2$ of 47 Tuc with adequate \textit{HST} photometry, and we finally identified 9 objects.

Summarizing, 9 BSSs have been excluded due to variability, 16 because they have been measured in too few bandpasses (or because only a few exposures were available per filter), and 6 because of the photometric quality selection.
Therefore, the final, quality-selected and variable-cleaned BSS sample consists of 22 stars (the dark blue dots in Figures~\ref{fig:FOV},~\ref{fig:CMDsel}~and~\ref{fig:vpd}), which still covers the whole magnitude and color ranges of the observed BSS sequence, thus allowing us to study this population in its entirety. 18 of the 22 BSSs from the clean sample have a PM measurement, that allowed us to definitely confirm that they are cluster members.

\subsection{Spectral Energy Distribution fitting}\label{ss:sedfitting}

We first corrected the observed magnitudes for reddening, adopting the following relation:

\begin{equation}\label{eq:2}
    m_{f,\mathrm{corr}} = m_{f,\mathrm{obs}} - c_{f}~\mathrm{R}_V~\mathrm{E}(B-V)
\end{equation}
where, for each bandpass $f$, $m_{f,\mathrm{corr}} $ is the reddening-corrected magnitude; $m_{f,\mathrm{obs}} $ is the original observed magnitude; $c_{f}=A_{\lambda}/A_{V}$ is the extinction law (\citealt{cardelli89}); $\mathrm{R}_V=3.12$ is the extinction coefficient; $\mathrm{E}(B-V)=0.04$ is the reddening value for 47~Tuc (\citealt{h96}).

We then constructed model SEDs as follows. We produced a grid of synthetic spectra with temperature and surface gravity ranges appropriate for BSSs: $5000 K<T_{\mathrm{eff}}<10000 K$ with a step of $100 K$; $3<\log(g)<5$ with a step of $0.1$. 
All the synthetic spectra were calculated with the software {\tt SYNTHE} \citep{sbordone04,kurucz05}. For each point of the grid, a one-dimensional, plane-parallel, LTE model atmosphere has been calculated with the code {\tt ATLAS9} \citep{kurucz05}, adopting [Fe/H]=--0.70 dex and an $\alpha$-enhanced chemical mixture\footnote{We tested if a small [Fe/H] variation ($\sim0.1$ dex) or a solar-scaled (instead of $\alpha$-enhanced) chemical mixture could significantly affect our results, and we found that the impact on the derived synthetic magnitudes is negligible.} \citep{dotter10}. The spectral synthesis was performed in the wavelength range between 1000 and 10000 \AA, including all the atomic and molecular lines available in the Kurucz/Castelli database\footnote{\url{http://wwwuser.oats.inaf.it/castelli/linelists.html}}, with the exclusion of TiO lines that are negligible for the investigated range of stellar parameters. Finally, each spectrum has been convolved with a Gaussian profile in order to obtain a spectral resolution of 1000.
The flux of synthetic spectra $F(\nu)$ is in units of $\mathrm{erg~cm^{-2}~s^{-1}~Hz^{-1}~sr^{-1}}$. We converted it into apparent flux $f(\lambda)$ (in \texttt{flam} units, i.e.: $\mathrm{erg~cm^{-2}~s^{-1}~}$\AA$^{-1}$) as follows:

\begin{equation}\label{eq:3}
    f_{\lambda}= \frac{4 \pi c}{\lambda^2} \Bigl(\frac{R}{D}\Bigr)^2 F(\nu)
\end{equation}
where $c=3\times10^{10}~\mathrm{cm~s}^{-1}$ is the speed of light; $R$ is the star radius (defined by a grid in the range: $0.1 R_{\odot}<R<4.5 R_{\odot}$, with a step of $0.01 R_{\odot}$); D is the cluster distance.
We converted the apparent fluxes to synthetic, apparent magnitudes $m_{f,\mathrm{syn}}$ (in the \texttt{VEGAMAG} system, to match our observed, calibrated magnitudes), by convolution with the filter throughputs, using the \texttt{pysynphot} package (\citealt{pysynphotref}).

We then directly compared observed and model SEDs. It is important to note that our model grid consists only of single star models, so we assume that all the BSSs in our sample are single stars. 
This assumption should be reasonably safe, since we have already excluded from the considered sample all the BSSs known or suspected to be eclipsing variables (see Sect.~\ref{ss:variab}). Moreover, the presence of BSSs with a degenerate companion (i.e., a WD) should not affect our results, since the WD emission is expected to be too hot and faint to significantly affect even the flux measured with the bluest filter.

\begin{figure*}[!t]
\centering
\includegraphics[width=.78\textwidth]{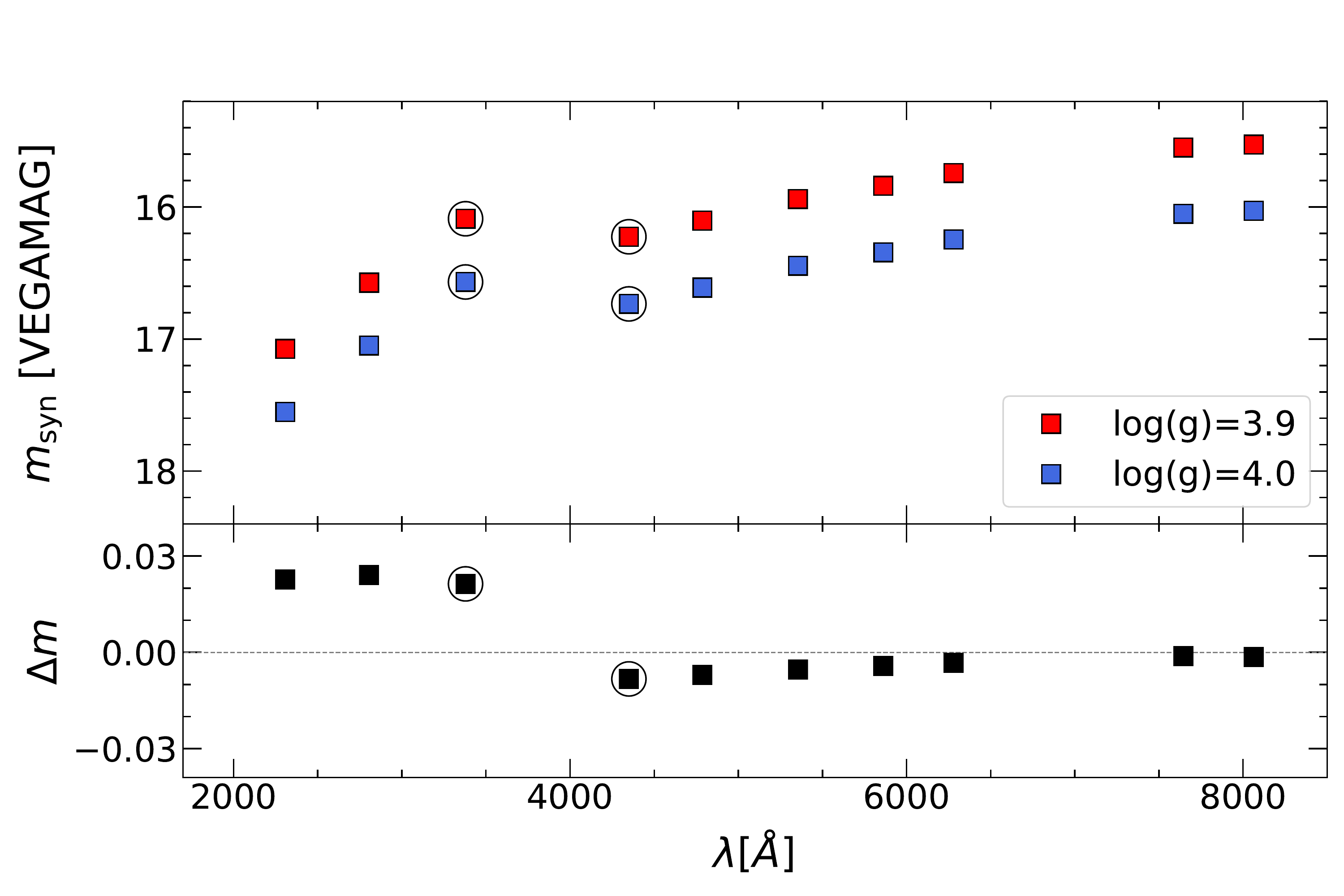}
\caption{Top panel: Synthetic magnitudes for two values of $\log(g)$, differing only by 0.1~dex (see legend), while temperature and radius are fixed ($T_{\mathrm{eff}}=7000~K, R=2.5~R_{\odot}$, respectively). The $\log(g)=4.0$ model is shifted by 0.5 mag fainter for clarity.
Bottom panel: the black squares corresponds to the difference between the two models plotted in the top panel. As it can be seen, although the difference between the two models is quite small (of the order of a few hundredths of magnitude), it is positive at wavelengths shorter than $\simeq 4000$ \AA, and negative redwards, changing sign abruptly. The quantity $\Delta m_{UB,\mathrm{syn}}=m_{\mathrm{F330W,syn}}-m_{\mathrm{F435W,syn}}$, defined in Sect.~\ref{ss:sedfitting}, has been introduced to maximise the sensitivity of the fit to this spectral region, where the dependence of the models on surface gravity is higher. $m_{\mathrm{F330W,syn}}$ and $m_{\mathrm{F435W,syn}}$ are, respectively, the third and fourth squares from the left, circled for clarity.}
\label{fig:MAGSYN}
\end{figure*}

We performed the fit using a Markov chain Monte Carlo (MCMC) approach, based on the \texttt{emcee} algorithm (\citealt{foremanmackey13}). We thus obtained the posterior probability distribution function (PDF) for each parameter of the fit ($T_{\mathrm{eff}}$, $\log(g)$, $R$) and subsequently derived the posterior PDF also for mass ($M$) and luminosity ($L$) through the following equations:

\begin{equation}\label{eq:4}
    g=\frac{GM}{R^2}; \ \ \  L=4\pi R^2 \sigma T_{\mathrm{eff}}^4
\end{equation}
where $G=6.67 \times 10^{-8}~\mathrm{g^{-1}~cm^3~s^{-2}}$ is the gravitational constant, and $\sigma=5.7 \times 10^{-5}~\mathrm{erg~cm^{-2}~s^{-1}~K^{-4}}$ is the Stefan-Boltzmann constant.
Since the priors we assumed are uniform, the posterior PDFs are proportional to the likelihood $L=\exp(-\chi^2/2)$.
We computed $\chi^2$ as:

\begin{equation}\label{eq:chi}
    \chi^2=\sum_{f}{\Bigl(\frac{\Delta m}{\sigma_{f,\mathrm{obs}}}\Bigr)^2} + \Bigl(\frac{\Delta m_{UB,\mathrm{corr}} - \Delta m_{UB,\mathrm{syn}}}{\sigma_{UB,\mathrm{obs}}}\Bigr)^2
\end{equation}
where the sum is performed over the 10 bandpasses used to construct the SEDs; $\Delta m = m_{f,\mathrm{corr}} - m_{f,\mathrm{syn}}$ is the difference between the observed, dereddened magnitudes and the synthetic ones; $\sigma_{f,\mathrm{obs}}$ is the error associated with the observed magnitudes (see Sect.~\ref{ss:2.1}). $\Delta m_{UB,\mathrm{corr}}=m_{\mathrm{F330W,corr}}-m_{\mathrm{F435W,corr}}$ is the observed magnitude difference (i.e., color) between the observed, dereddened magnitudes in the F330W and F435W bandpasses (roughly corresponding to Johnson $U$ and $B$ filters), $\Delta m_{UB,\mathrm{syn}}=m_{\mathrm{F330W,syn}}-m_{\mathrm{F435W,syn}}$ is the equivalent quantity for synthetic magnitudes, while $\sigma_{UB,\mathrm{obs}}$ is the error associated to $\Delta m_{UB,\mathrm{corr}}$, obtained by adding in quadrature $\sigma_{\mathrm{F330W,obs}}$ and $\sigma_{\mathrm{F435W,obs}}$, i.e., the errors associated to the observed magnitudes in these two bandpasses. The aim of this last term of the $\chi^2$ is to increase the sensitivity of the fit to surface gravity. As can be seen in Figure~\ref{fig:MAGSYN}, surface gravity has quite a weak impact on the SEDs. 
However, the spectral region sampled by the F330W and F435W filters (i.e., around the Balmer jump region; circled squares in Figure~\ref{fig:MAGSYN}) seems instead to be sensitive to the adopted gravity. 
Increasing the fit sensitivity to surface gravity is particularly important for two reasons: first, the surface gravity grid is defined in logarithmic units; therefore, small uncertainties in $\log(g)$ translate into significant variations of the surface gravity. Second, mass is directly proportional to surface gravity, therefore the goodness of the fit on surface gravity directly influences the accuracy of the mass determination.

\begin{table*}[!t]
\caption{Best fit parameters for the clean BSS sample.\label{tab:results}}
\centering
{
\begin{tabular}{llllrl}
\hline\hline
BSS ID & $T \ \mathrm{[K]}$ & $\log(g)$ & $R \ \mathrm{[R_{\odot}]}$ & $L \ \mathrm{[L_{\odot}]}$ & $M \ \mathrm{[M_{\odot}]}$ \\
\hline
\hline
BSS1  & $7600^{+80 }_{-90 }$ & $3.70^{+0.11}_{-0.13}$ & $3.51^{+0.11}_{-0.09}$ & $35.9^{+0.8}_{-0.8}$ & $2.31^{+0.64}_{-0.59}$ \\
BSS2  & $7900^{+90 }_{-90 }$ & $3.63^{+0.09}_{-0.07}$ & $2.99^{+0.08}_{-0.08}$ & $30.7^{+0.7}_{-0.7}$ & $1.39^{+0.35}_{-0.22}$ \\
BSS4  & $7600^{+110}_{-110}$ & $3.79^{+0.12}_{-0.11}$ & $2.92^{+0.09}_{-0.09}$ & $24.7^{+0.6}_{-0.6}$ & $1.93^{+0.57}_{-0.45}$ \\
BSS10 & $7800^{+130}_{-130}$ & $4.08^{+0.11}_{-0.11}$ & $1.95^{+0.06}_{-0.07}$ & $12.5^{+0.3}_{-0.3}$ & $1.68^{+0.56}_{-0.40}$ \\
BSS12 & $7100^{+90 }_{-50 }$ & $3.72^{+0.10}_{-0.10}$ & $2.67^{+0.04}_{-0.08}$ & $15.9^{+0.4}_{-0.3}$ & $1.36^{+0.33}_{-0.31}$ \\
BSS14 & $7200^{+90 }_{-90 }$ & $3.90^{+0.11}_{-0.13}$ & $2.39^{+0.07}_{-0.06}$ & $13.8^{+0.4}_{-0.4}$ & $1.66^{+0.54}_{-0.43}$ \\
BSS17 & $7200^{+50 }_{-80 }$ & $3.74^{+0.09}_{-0.08}$ & $2.43^{+0.08}_{-0.03}$ & $13.9^{+0.3}_{-0.3}$ & $1.19^{+0.27}_{-0.22}$ \\
BSS18 & $7400^{+70 }_{-80 }$ & $3.90^{+0.10}_{-0.09}$ & $2.17^{+0.06}_{-0.03}$ & $12.3^{+0.3}_{-0.3}$ & $1.39^{+0.37}_{-0.25}$ \\
BSS20 & $6800^{+70 }_{-60 }$ & $3.82^{+0.11}_{-0.10}$ & $2.71^{+0.03}_{-0.09}$ & $13.8^{+0.4}_{-0.3}$ & $1.75^{+0.53}_{-0.38}$ \\
BSS22 & $6400^{+50 }_{-60 }$ & $3.41^{+0.11}_{-0.10}$ & $3.72^{+0.13}_{-0.04}$ & $20.4^{+0.6}_{-0.5}$ & $1.29^{+0.38}_{-0.28}$ \\
BSS23 & $6700^{+50 }_{-70 }$ & $3.75^{+0.12}_{-0.12}$ & $2.75^{+0.10}_{-0.04}$ & $13.4^{+0.3}_{-0.3}$ & $1.57^{+0.60}_{-0.37}$ \\
BSS24 & $7100^{+50 }_{-100}$ & $4.18^{+0.12}_{-0.10}$ & $1.92^{+0.06}_{-0.03}$ & $8.2 ^{+0.2}_{-0.2}$ & $2.10^{+0.66}_{-0.45}$ \\ 
BSS25 & $6900^{+70 }_{-50 }$ & $4.00^{+0.12}_{-0.11}$ & $2.20^{+0.04}_{-0.06}$ & $9.6 ^{+0.2}_{-0.2}$ & $1.76^{+0.56}_{-0.42}$ \\ 
BSS27 & $6200^{+60 }_{-50 }$ & $3.15^{+0.15}_{-0.11}$ & $3.82^{+0.06}_{-0.13}$ & $18.9^{+0.6}_{-0.5}$ & $0.76^{+0.31}_{-0.16}$ \\
BSS28 & $6900^{+60 }_{-60 }$ & $4.02^{+0.10}_{-0.11}$ & $2.00^{+0.06}_{-0.03}$ & $7.9 ^{+0.2}_{-0.2}$ & $1.56^{+0.42}_{-0.35}$ \\ 
BSS30 & $6700^{+90 }_{-40 }$ & $3.83^{+0.11}_{-0.10}$ & $2.24^{+0.03}_{-0.06}$ & $8.8 ^{+0.2}_{-0.2}$ & $1.22^{+0.34}_{-0.25}$ \\ 
BSS31 & $6700^{+80 }_{-60 }$ & $3.72^{+0.14}_{-0.14}$ & $2.21^{+0.03}_{-0.07}$ & $8.7 ^{+0.2}_{-0.2}$ & $0.92^{+0.37}_{-0.27}$ \\ 
BSS32 & $6400^{+60 }_{-50 }$ & $3.72^{+0.18}_{-0.19}$ & $2.79^{+0.06}_{-0.06}$ & $11.4^{+0.4}_{-0.3}$ & $1.49^{+0.77}_{-0.54}$ \\
BSS42 & $6600^{+70 }_{-50 }$ & $3.93^{+0.11}_{-0.11}$ & $1.83^{+0.02}_{-0.06}$ & $5.6 ^{+0.2}_{-0.2}$ & $1.03^{+0.30}_{-0.25}$ \\
BSS43 & $6200^{+90 }_{-50 }$ & $3.39^{+0.20}_{-0.20}$ & $2.65^{+0.05}_{-0.09}$ & $9.1 ^{+0.3}_{-0.3}$ & $0.63^{+0.33}_{-0.26}$ \\
BSS49 & $6600^{+60 }_{-80 }$ & $3.87^{+0.17}_{-0.16}$ & $1.53^{+0.05}_{-0.02}$ & $3.9 ^{+0.1}_{-0.1}$ & $0.64^{+0.33}_{-0.20}$ \\
BSS52 & $6300^{+80 }_{-60 }$ & $3.84^{+0.18}_{-0.15}$ & $1.78^{+0.03}_{-0.06}$ & $4.4 ^{+0.1}_{-0.1}$ & $0.79^{+0.45}_{-0.24}$ \\

\hline
\end{tabular}}
\end{table*}

\begin{figure}[!t]
\centering
\includegraphics[width=\textwidth]{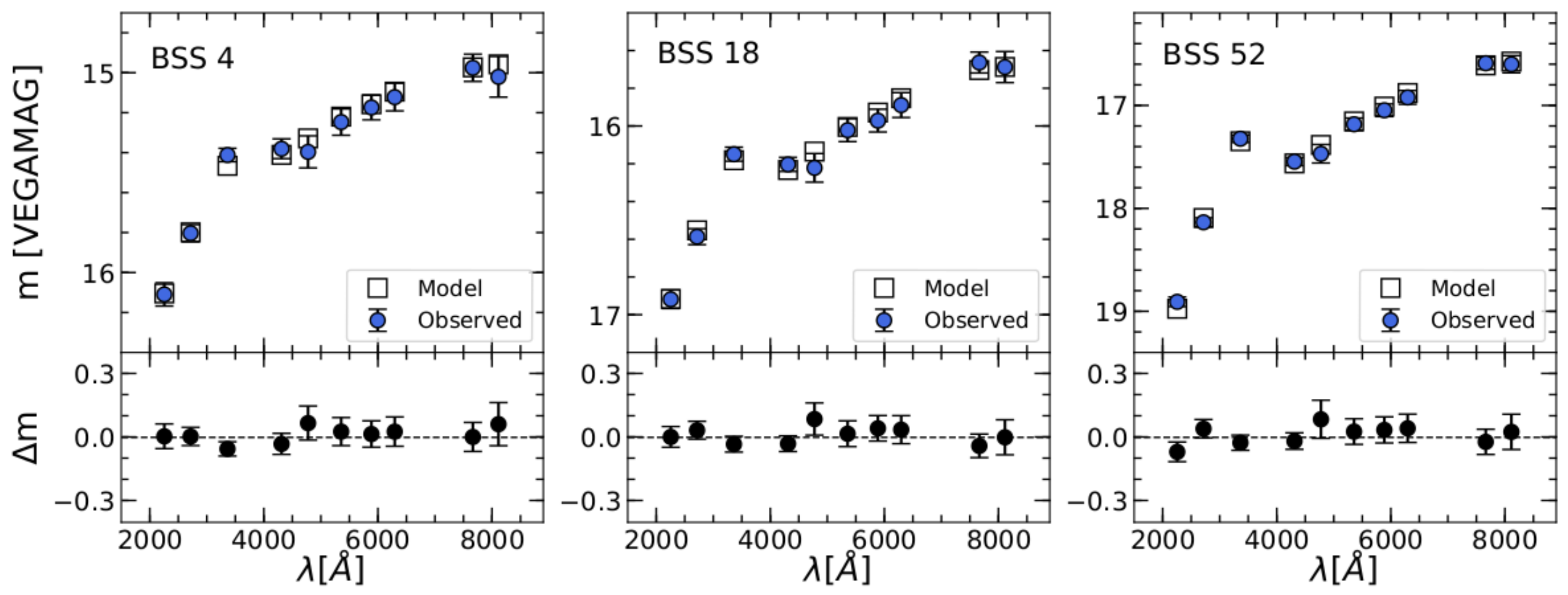}
\caption{Top panels: observed SEDs of three BSSs from our sample, namely BSS4, BSS18, BSS52 (blue circles), overplotted to the corresponding best fit model (empty squares; see Table~\ref{tab:results}). Bottom panels: residuals between the observed SED and the best-fit model.}
\label{fig:sedfit}
\end{figure}

The results of the fitting procedure are listed in Table~\ref{tab:results}. The best fit values for each parameter correspond to the PDF median, while the reported uncertainties correspond to the $68\%$ confidence interval. In Figure~\ref{fig:sedfit} we show the SEDs of a bright, an intermediate-magnitude and a faint BSSs, namely BSS4, BSS18, BSS52, overplotted to the corresponding best-fit model. In the lower panels we show the residuals between the observed SED and the best-fit model. It can be seen that, in any case, the residuals are small and that observed SEDs and models are in good agreement within the errors.

\section{Discussion}\label{s:4}

\begin{figure}[!t]
\centering
\includegraphics[height=.4\textheight]{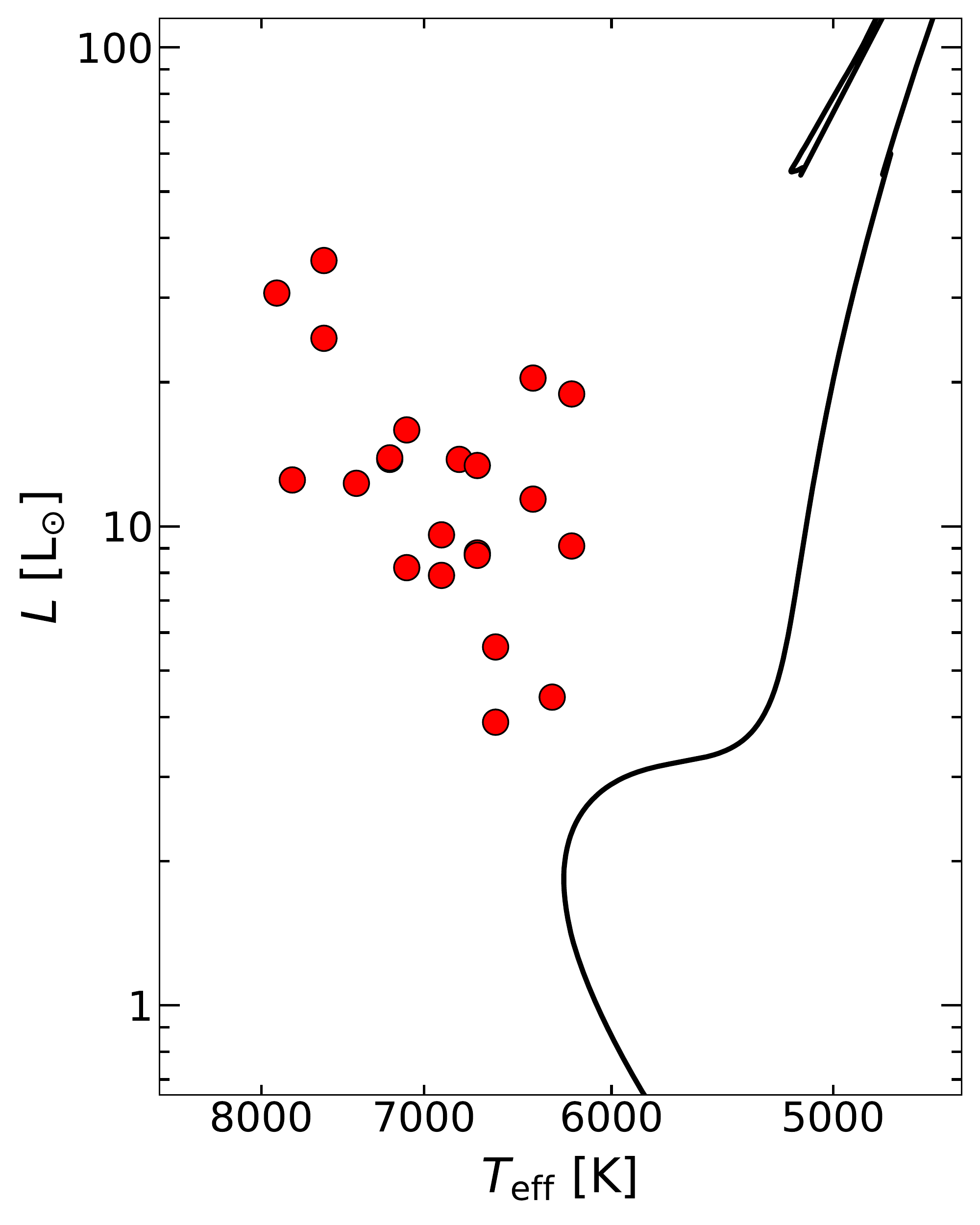}
\caption{Position of the studied BSSs (red circles) in the HR diagram. The black solid line is a 12~Gyr BaSTI isochrone (\citealt{pietrinferni04, hidalgo18}), plotted as a reference to trace the normal TO, SGB and RGB sequences of the cluster.}
\label{fig:diagHR}

\centering
\includegraphics[width=\textwidth]{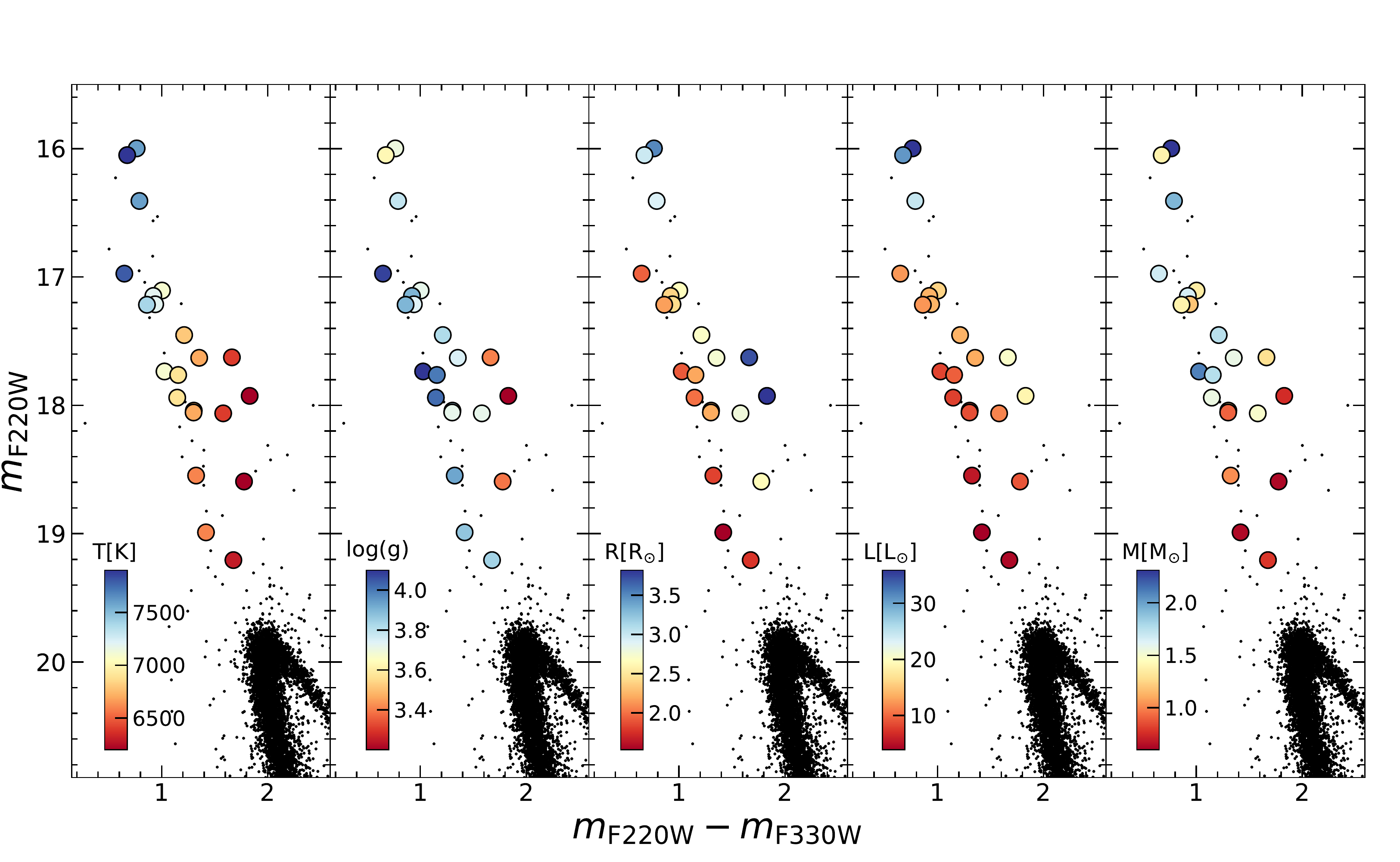}
\caption{Distributions of the best-fit parameters obtained for the clean BSS sample, shown in the CMD. From left to right: temperature, surface gravity (logarithmic units), radius, bolometric luminosity, mass. Units and colorbars are marked inside each panel.}\label{fig:RESULTS}
\end{figure}

In Figure~\ref{fig:diagHR} we plot the luminosity and temperature of the selected BSSs in the ($\log L - \log T_{\mathrm{eff}}$) Hertzsprung-Russel (HR) diagram. As a reference, we also plot a 12~Gyr BaSTI isochrone\footnote{\url{http://basti-iac.oa-abruzzo.inaf.it/index.html}} (\citealt{pietrinferni04, hidalgo18}, solid black line) to highlight the TO, sub-giant branch (SGB), red giant branch (RGB) and horizontal branch (HB) loci of ``normal'' stars of the cluster. As expected, the values of temperature and luminosity derived from the SED fitting make BSSs standing clearly outside the standard evolutionary loci, defining a sequence along the extrapolation of the cluster MS, with luminosities ranging from $\sim3$ to $\sim30~L_{\odot}$ and temperatures between $\sim6000$ to $\sim8000$~K.

\begin{figure}[!t]
\centering
\includegraphics[width=.63\textwidth]{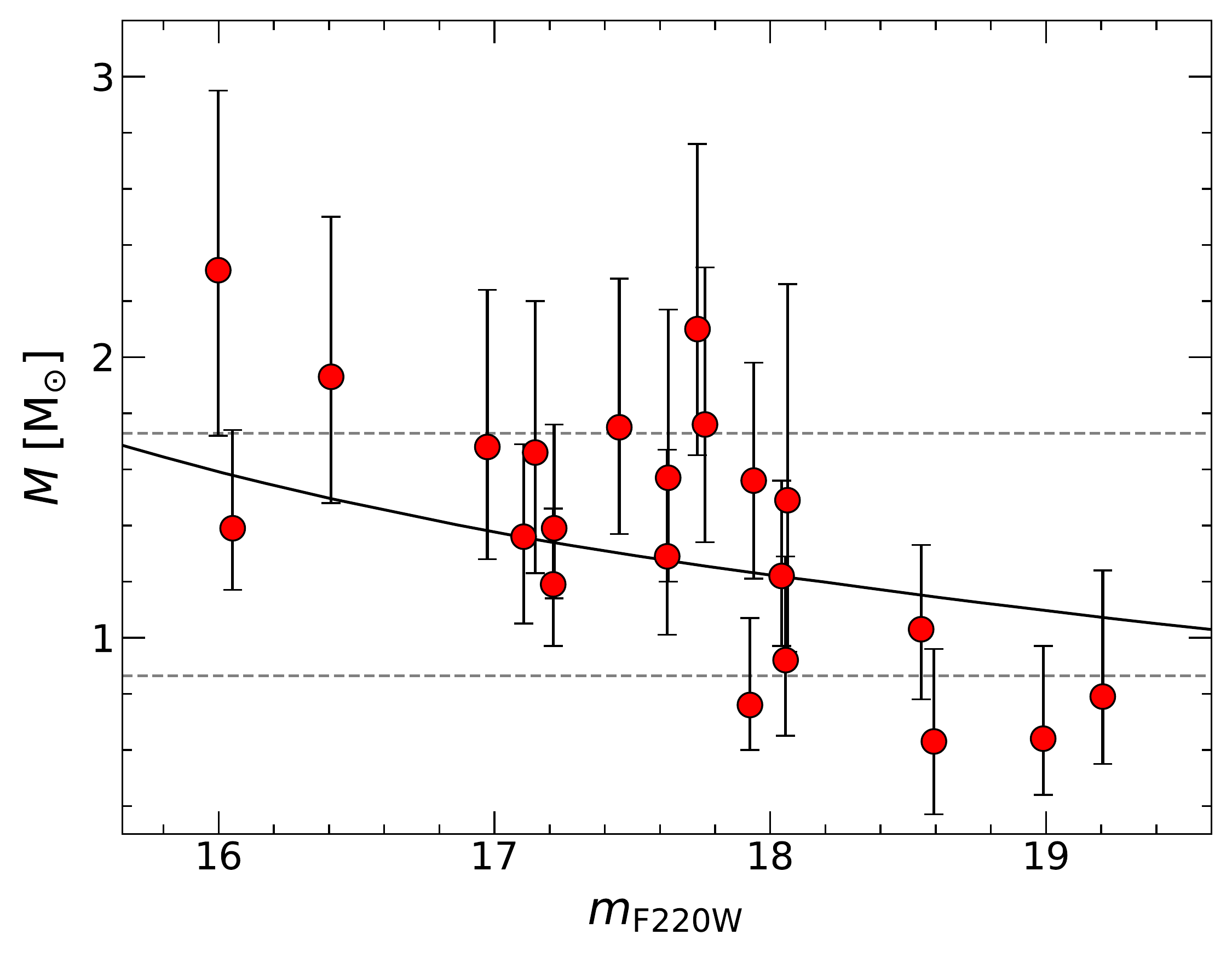}
\caption{Resulting masses, with uncertainties corresponding to the $68\%$ confidence interval, as a function of F220W magnitude (red points). The black solid line is a 100~Myr BaSTI isochrone (\citealt{pietrinferni04, hidalgo18}) with [Fe/H]=-0.70, plotted as a reference. The two horizontal, dashed lines corresponds to $M_{\mathrm{TO}}$ and $2M_{\mathrm{TO}}$ ($M_{\mathrm{TO}}\sim0.86~M_{\odot}$, from the 12~Gyr BaSTI isochrone reported in Figure~\ref{fig:diagHR}). }\label{fig:MASSMAG}
\end{figure}

In Figure~\ref{fig:RESULTS} we show the distribution of the analyzed BSSs in the UV CMD, in which each BSS is marked with a color code that quantifies the value of the parameters derived from the SED analysis (see labels). 
As can be seen, the resulting parameters vary according to theoretical predictions. In the first panel, temperature decreases as BSS color increases. In the second and third panels respectively, surface gravity decreases and radius increases moving away from the zero age main sequence (ZAMS). In the fourth panel, bolometric luminosity increases as a function of the magnitude, but there is also an expected dependence on the color, since at fixed UV magnitudes, the reddest stars have lower temperatures, thus they are brighter at longer wavelengths (i.e., they have larger bolometric corrections): hence their bolometric luminosity is larger.

Regarding BSS masses, the distribution shown in the rightmost panel of Figure~\ref{fig:RESULTS} suggests the presence of a mass succession along the BSS sequence, with lower masses at the faint end and higher masses at the bright end of the sequence.  
We reiterate, however, that mass errors are quite large (see column 5 of Table~\ref{tab:results}, and Figure~\ref{fig:MASSMAG}) due to the lack of very strong surface gravity tracers in the SEDs, as explained in the previous section. In spite of this, a mass sequence is also confirmed by the plot in Figure~\ref{fig:MASSMAG}, where the BSS masses are plotted as a function of the F220W magnitude.  
Moreover, the mass distribution seems qualitatively in agreement with that predicted by theoretical ZAMS models: the black line in Figure~\ref{fig:MASSMAG} is a 100~Myr BaSTI isochrone (\citealt{pietrinferni04, hidalgo18}) with [Fe/H]$=-0.70$, plotted as a reference for the ZAMS of the cluster. 
In order to test the statistical significance of the detected BSS mass-magnitude relation, we measured the Spearman and Pearson correlation coefficients. We obtained $\rho~=~-0.64$ and $r~=~-0.70$ respectively, supporting the presence of an anti-correlation between mass and magnitude.
Note that the mass-magnitude relation is also visible by using different bandpasses, but the adoption of UV filters tends to maximize the magnitude extension of the BSS sequence.

As shown in Figure~\ref{fig:MASSMAG}, a few BSSs (specifically BSS1, BSS4, BSS20, BSS24, BSS25) in our sample have masses larger than $2M_{\mathrm{TO}}$ (we assumed $M_{\mathrm{TO}}\sim0.86~M_{\odot}$, from the same 12~Gyr BaSTI isochrone reported in Figure~\ref{fig:diagHR}). The presence of BSSs with twice the MS-TO mass would imply a formation mechanism that involves at least three stars (e.g., \citealt{knigge06,knigge08}). However, within the uncertainties, these stars are still compatible with a mass lower than this threshold value, therefore they are just candidate supermassive-BSSs. It would be interesting to measure the masses of these stars with other direct methods, e.g., spectroscopically (see \citealt{ferraro16}), in order to confirm or reject this hypothesis. 
The masses derived for 4 BSSs (specifically BSS27, BSS43, BSS49, BSS52), turn out to be lower than $M_{\mathrm{TO}}$. These values are, however, still compatible with a mass larger than the TO mass within the errors. Most of them are (with the possible exception of BSS27) low-mass faint BSSs: they are clearly distinguishable from the MS only in an UV-CMD, as the one used in this work, while in a classical, optical diagram they do not appear to be significantly different from MS-TO stars (see \citealt{raso17}).

\begin{figure}[!t]
\centering
\includegraphics[height=.57\textheight]{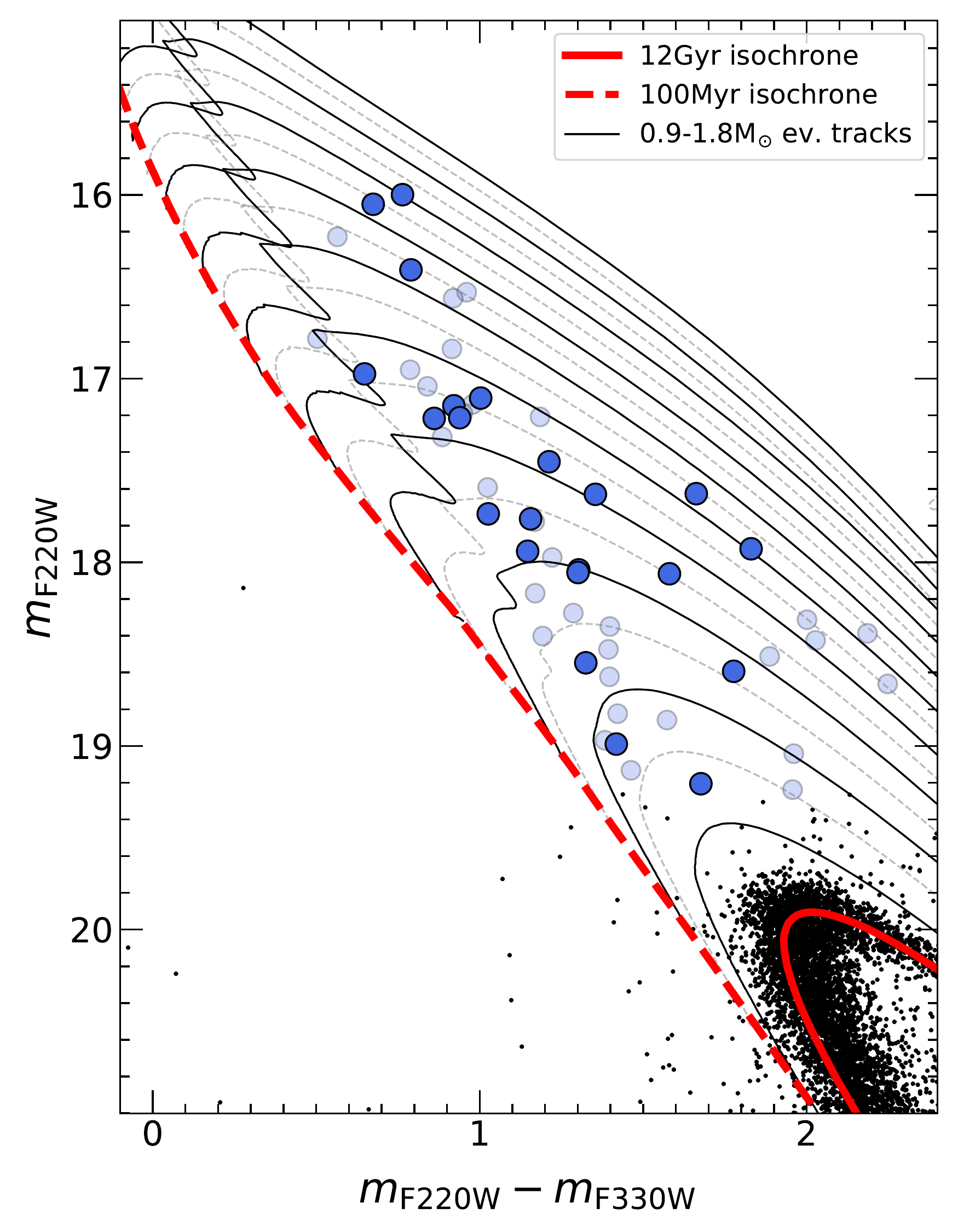}
\caption{UV CMD with the BSS sample higlighted (blue circles; the dark blue ones represent the clean sample). The evolutionary tracks used to estimate the BSS mass ($M_{\mathrm{TR}}$, see text for details) are also plotted. They span the range $0.9-1.8~M_{\odot}$, with a step of $0.05~M_{\odot}$ (alternatively plotted as solid black lines and grey dashed lines for clarity). Two isochrones, of 12~Gyr and 100~Myr, are superposed for reference, respectively as solid and dashed red lines.}\label{fig:cmdTRACKS}

\centering
\includegraphics[height=.32\textheight]{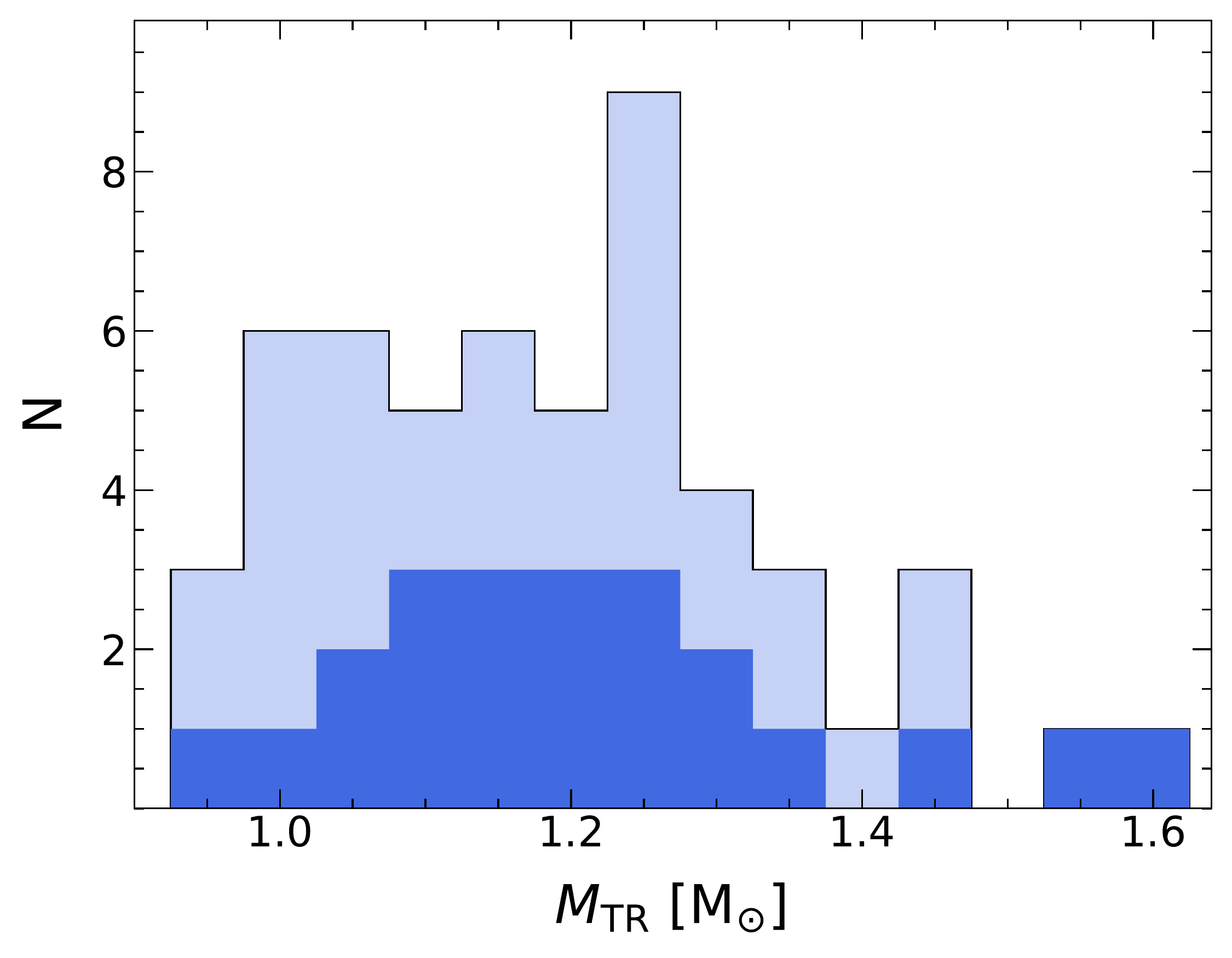}
\caption{Light blue histogram: mass distribution, obtained from the evolutionary tracks, for the entire sample of 53 BSSs. Dark blue histogram: mass distribution, obtained from the evolutionary tracks, for the clean sample of 22 BSSs.}\label{fig:MASSfunc}
\end{figure}

In general, a comparison with evolutionary tracks can provide a rough estimate of the BSS mass. Here we used a set of isochrones and evolutionary tracks (in the range $0.9-1.8 M_{\odot}$) from the BaSTI  model library. The theoretical models have been colored  by convolving a grid of suitable \citet{kurucz93} stellar spectra of appropriate metallicity with the transmission curves of the used ACS/HRC filters. Thus, for each given stellar temperature and gravity, both the color and the bolometric corrections in the \texttt{VEGAMAG} system have been computed. The set of evolutionary tracks (at [Fe/H]=-0.7) in the range $0.9-1.8 M_{\odot}$ are shown in Figure~\ref{fig:cmdTRACKS}, over-plotted to the ($m_{\mathrm{F220W}}$ vs. $m_{\mathrm{F220W}}-m_{\mathrm{F330W}}$) CMD: a distance modulus of $(m-M)_0=13.3$ and a color excess E($B-V$)=0.04 have been adopted. Small offsets (of the order of a few 0.01) in color and magnitude have been applied to the evolutionary models in order to allow the 12~Gyr isochrone to perfectly match the MS-TO. 
The evolutionary tracks at steps of $0.05~M_{\odot}$ shown in Figure~\ref{fig:cmdTRACKS} represent the reference ``pillars'' for the interpolation procedure that allowed us to estimate the mass of each BSS. The derived mass distribution for the entire sample of 53 BSSs is plotted in Figure~\ref{fig:MASSfunc}.
An average mass of $1.2~M_{\odot}$, in agreement with other mass determination in the literature (see \citealt{ferraro06, lanzoni07b, fiorentino14}), is obtained from this sample. 

\begin{figure}[!t]
\centering
\includegraphics[height=.45\textheight]{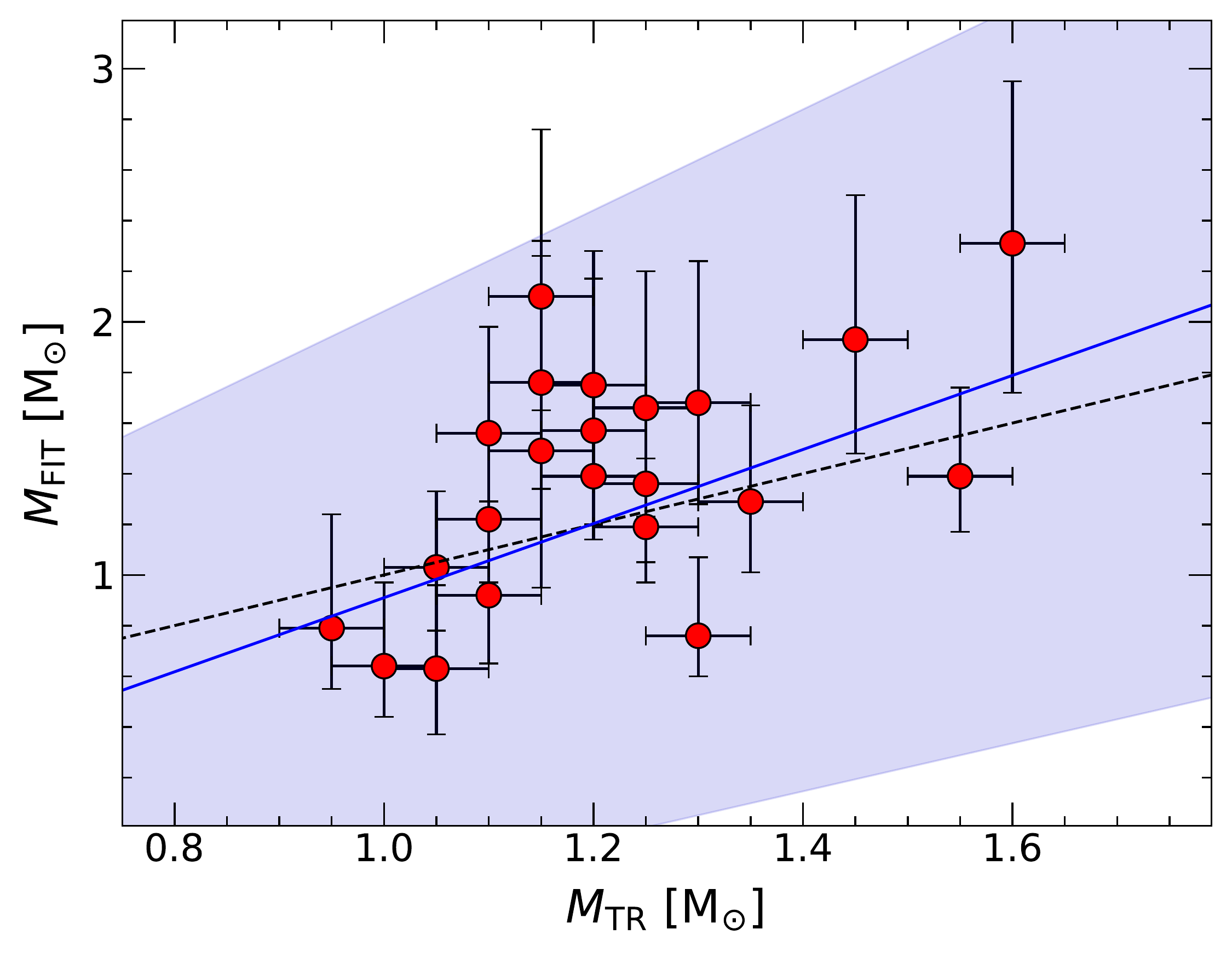}
\caption{Red points: masses resulting from the SED-fitting method, with uncertainties corresponding to the $68\%$ confidence interval, as a function of the masses estimated from evolutionary tracks (see text for details; the $M_{\mathrm{TR}}$ uncertainties are equal to 0.05~$\mathrm{M_{\odot}}$, i.e., the step of the evolutionary grid used to measure $M_{\mathrm{TR}}$ itself). The black, dashed line is the bisector. The blue solid line is the best-fit straight line obtained with a maximum likelihood approach, considering both the y- and x-axis uncertainties (with slope $a=1.46^{+0.53}_{-0.52}$ and intercept $b=-0.55^{+0.60}_{-0.63}$). The blue shaded area represents the 68\% interval around the best-fit relation.}\label{fig:MASSTR}
\end{figure}

Note that here we have the possibility of directly comparing the BSS masses obtained from the SED-fitting ($M_{\mathrm{FIT}}$) with those obtained from evolutionary tracks ($M_{\mathrm{TR}}$), for the sub-sample of 22 well-measured BSSs.
We report the result of the comparison between $M_{\mathrm{FIT}}$ and $M_{\mathrm{TR}}$ in Figure~\ref{fig:MASSTR}. 
We fitted the points with a straight line using a maximum likelihood approach, equivalent to the one described in Sect.~\ref{ss:sedfitting}, considering both the y- and x-axis uncertainties. 
We obtained a slope $a=1.46^{+0.53}_{-0.52}$ and an intercept $b=-0.55^{+0.60}_{-0.63}$ (the blue solid line in Figure~\ref{fig:MASSTR}). In the figure we also plotted, as a reference, the bisector line (black dashed line) i.e., a straight line with slope $a=1$ and intercept $b=0$, which represents the full correspondence between the $M_{\mathrm{FIT}}$ and $M_{\mathrm{TR}}$ values. 
As can be seen, the best fit relation turns out to be in reasonable agreement with the bisector. Therefore, we can conclude that, when a direct BSS mass estimate cannot be made, the evolutionary track method can be used to provide a reasonable first-guess estimate of the BSS mass.

\subsection{Comparison with previous mass estimates}\label{ss:compmass}

\begin{figure}[!t]
\centering
\includegraphics[width=.95\textwidth]{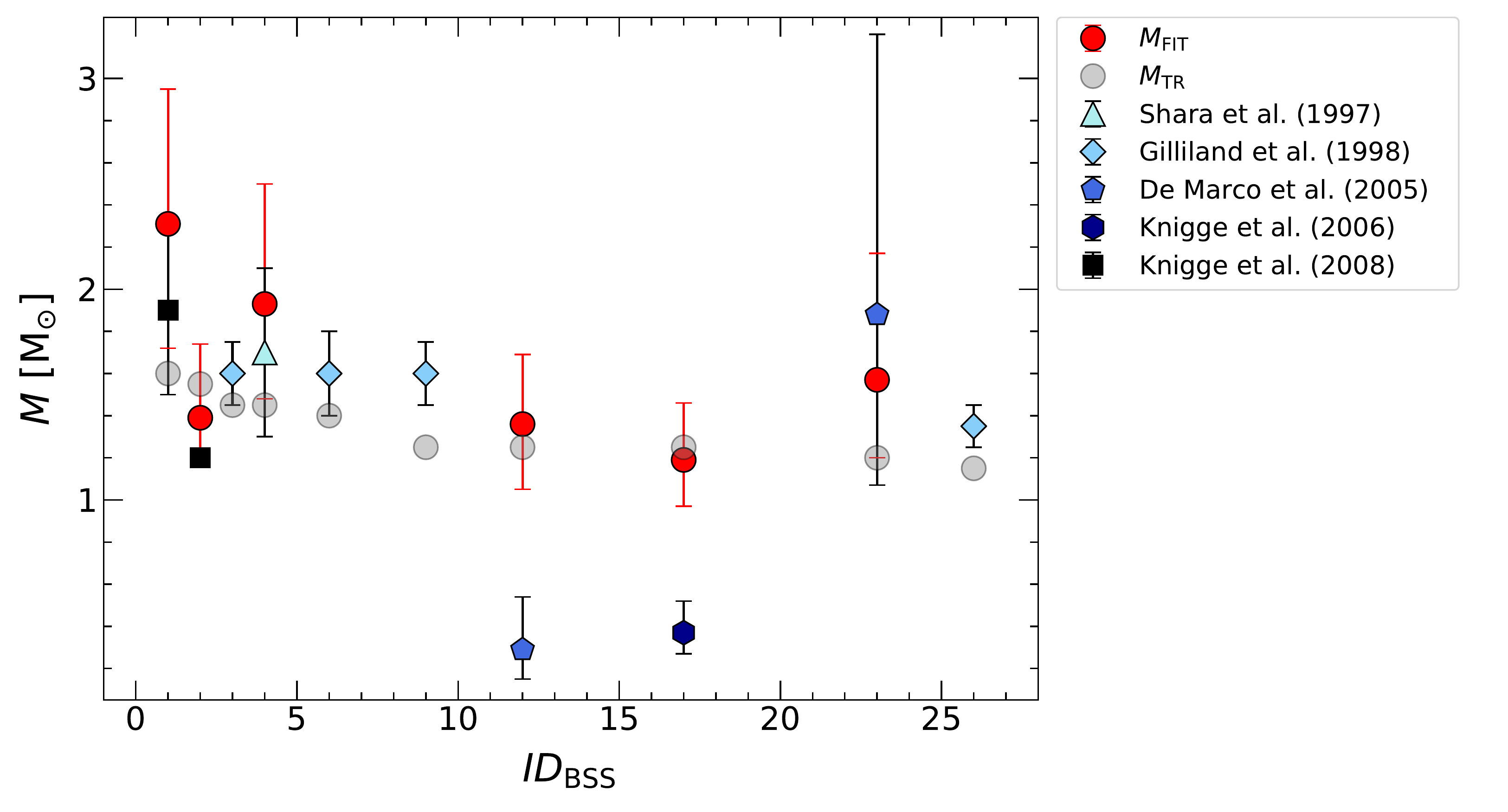}
\caption{BSS masses obtained from our SED fitting method (red circles) and from the comparison with evolutionary tracks (grey circles), plotted against the BSS ID, compared to the values quoted in the literature for the same objects (points with different shades of blue, see legend).}\label{fig:MASScomp}
\end{figure}

A few direct mass measurements of BSSs in the core of 47~Tuc are already available in the literature. In this section, we compare our results with previous estimates. The results of this comparison are summarised in Figure~\ref{fig:MASScomp}.

The first direct BSS mass measurement was indeed obtained on a 47~Tuc BSS by \citet{shara97}, comparing an \textit{HST} spectrum (obtained with the Faint Object Spectrograph) with model spectra, and finding $M=1.7\pm0.4~M_{\odot}$ (light blue triangle in Figure~\ref{fig:MASScomp}). The BSS they studied corresponds to our BSS4. The mass we obtain for this object is $M=1.93^{+0.57}_{-0.45}~M_{\odot}$ (see Table~\ref{tab:results}), which is fully compatible, within the errors, with the value obtained spectroscopically by \citet{shara97}.

\citet{gilliland98} measured the masses of four variable BSSs in 47~Tuc on the basis of their pulsations (specifically V2, V14, V15, V16). They found masses equal to: $1.6\pm0.2~\mathrm{M_{\odot},} \ 1.35\pm0.1~\mathrm{M_{\odot},} \ 1.6\pm0.15~\mathrm{M_{\odot},} \ 1.6\pm0.15~\mathrm{M_{\odot}}$, respectively (light blue diamonds in Figure~\ref{fig:MASScomp}).
We indeed detected in our FOV the four BSSs studied in \citet{gilliland98}, which correspond to BSS6, BSS26, BSS3 and BSS9, respectively. We did not estimate their mass through SED fitting, since we excluded from this study all the known or suspected variables (see Sect.~\ref{ss:variab}).
However, we can compare the values reported in \citet{gilliland98} with the masses we obtained from evolutionary tracks. Specifically, our $M_{\mathrm{TR}}$ values for these four objects are: 1.40~$\mathrm{M_{\odot}}$, 1.15~$\mathrm{M_{\odot}}$, 1.45~$\mathrm{M_{\odot}}$ and 1.25~$\mathrm{M_{\odot}}$. As can be seen, the values for BSS3 and BSS6 are compatible with the values from \citet{gilliland98}, while for BSS9 and BSS26 the masses from their work are slightly larger. However, estimating masses from evolutionary tracks could be less efficient for variable stars, since this kind of objects slightly changes position in the CMD depending on the phase of the variability.

\citet{demarco05} spectroscopically obtained masses and rotation rates of 55 stars (including 24 BSSs) in four GCs, including 47~Tuc. In particular, they analyzed 5 BSSs in 47~Tuc (see their Table~4). We detected all 5 of these objects in our sample, but only two of them pass our quality/variability conditions and therefore they have been analyzed in this work. Namely, their star NGC104-5 corresponds to our BSS12 and their NGC104-7 corresponds to our BSS23. They obtain masses equal to $0.29^{+0.25}_{-0.14}~M_{\odot}$ and $1.88^{+1.33}_{-0.81}~M_{\odot}$, respectively (blue pentagons in Figure~\ref{fig:MASScomp}), while we obtain $1.36^{+0.33}_{-0.31}~M_{\odot}$ and $1.57^{+0.60}_{-0.37}~M_{\odot}$, respectively. The two mass values for BSS23 are fully compatible within the errors, while those of BSS12 are not, with their mass estimate being significantly smaller than ours. The spectrum of this star has a high S/N ratio but is affected by blending (see Sect.~7.2 in \citealp{demarco05}). The authors also discuss a temperature inconsistency between the low- and the intermediate-resolution spectra (see their Sect.~11). These can be two possible explanations for the discrepancy between the two mass measurements.

\citet{knigge06,knigge08} obtained physical parameter estimates for a few BSSs in the core of 47~Tuc using almost the same dataset as ours (excluding the ACS/HRC F220W images and including the ACS/HRC F850LP and STIS F25QTZ images, and also FUV spectroscopy in the latter work) and the same technique, i.e., SED fitting, albeit using a least squares approach.

In \citet{knigge06}, they identify the star BSS7 (dark blue hexagon in Figure~\ref{fig:MASScomp}; nomenclature from \citealt{paresce91}; hereafter, to avoid confusion with our nomenclature, we will name this star K-BSS7) as the optical counterpart to the \textit{Chandra} X-ray source W31 (\citealt{grindlay01}). They found K-BSS7 to be variable, but with a very small amplitude ($A_{I}=0.0037$~mag). K-BSS7 corresponds to BSS17 in this work. Given its very small variability amplitude, BSS17 survived the selection criteria discussed in Sect.~\ref{ss:variab} and has been kept in our final sample. \citet{knigge06} fit the SED with both a single and a binary model, finding comparable results for the physical parameters of the BSS, although the presence of a MS secondary improved the quality of their fit. Irrespective of the presence of a companion, the physical parameters they found from the fits point towards quite low mass values ($M=0.34^{+0.15}_{-0.08}~M_{\odot}$ and $M=0.37^{+0.15}_{-0.10}~M_{\odot}$ for the single and binary models, respectively). They argue that such low mass values can be due to systematic uncertainties in the $\log(g)$ estimates due to uncertainties on, e.g., the cluster distance, reddening and metallicity, or that the low $\log(g)$ values can be due to the rapid rotation of the star. Unfortunately, without spectroscopy it is not possible to disentangle between the two scenarios. 
On the contrary, our mass estimate for BSS17 is $M=1.19^{+0.27}_{-0.22}~M_{\odot}$, larger than the TO mass and compatible with the value found for other BSSs of comparable magnitude. It is important to note that the cluster distance, reddening and metallicity we adopted are slightly different from theirs ($0.01$ kpc in distance, $-0.13$ dex in metallicity and $0.008$ in reddening), but these small differences can hardly explain the large discrepancy in the derived mass. The significant difference between these two mass values for BSS17 might be explained by the fact tha we use a MCMC approach to the SED fitting procedure, and we explicitly add a term in the $\chi^2$ computation (see Eq.~\ref{eq:chi}) to increase the sensitivity of the fit to the surface gravity.

\citet{knigge08} used FUV spectroscopy (obtained with the Space Telescope Imaging Spectrograph onboard \textit{HST}) to study 48 FUV-excess sources in the same FOV as this work, classifying them on the basis of their FUV-optical CMD and combining the FUV spectroscopy with UV-optical SEDs (constructed using the same photometric dataset used in \citealp{knigge06}, almost coincident with ours; see previous paragraph) to further study these sources and constrain their physical parameters. They found 8 BSSs in their sample, but they analyzed in detail, i.e., obtained physical parameter estimates, only for two of them (star 2 and star 999 using their nomenclature; shown as black squares in Figure~\ref{fig:MASScomp}). These two stars correspond to our BSS2 and BSS1, respectively. 
Regarding star 2 (BSS2), \citet{knigge08} detected a significant FUV excess from spectroscopy, which they associate to a WD companion. However, the photometric data can still be described by a single component since the WD emission is too blue to have a significant impact at those wavelengths. Therefore, regarding the BSS physical parameters, we can safely compare our result, obtained under the assumption of single stars (see Sect.~\ref{ss:sedfitting}), with theirs. They obtain a mass of $1.2~M_{\odot}$ (no uncertainties reported), consistent, within the errors, with our result for BSS2 ($M=1.39^{+0.35}_{-0.22}~M_{\odot}$).
Regarding star 999 (BSS1), they obtain $M=1.9\pm0.4~M_{\odot}$, fitting the broadband SED only, excluding the FUV spectrometry (see their Sect.~4.12 for a detailed explanation). This is also comparable to our measurement: $M=2.31^{+0.64}_{-0.59}~M_{\odot}$. Our BSS1 mass measurement seems to further confirm that this star, as already discussed in \citet{knigge08}, has a mass larger than $2M_{\mathrm{TO}}$. Unfortunately, both in \citet{knigge08} and in this work, the mass uncertainties are too large to definitively conclude that BSS1 must have had more than two progenitors.

\section{Conclusions}\label{s:5}

We used high angular resolution \textit{HST} ACS/HRC images to construct broadband BSS SEDs in the GC 47~Tuc. We obtained physical parameter estimates (temperature, gravity, radius, luminosity and mass) for 22 BSSs through SED fitting, using a MCMC approach. This is the first time that BSS masses have been obtained directly and consistently for such a large BSS sample within the same cluster.

The physical parameters we obtained are in good agreement with theoretical predictions, and in particular we find that BSSs in 47~Tuc define a mass sequence, with lower masses at fainter magnitudes and higher masses at brighter magnitudes.

We compare our SED-fit based mass measurements with estimates obtained from the BSS position in the CMD and a grid of evolutionary tracks, finding consistent results. Hence, when direct BSS mass estimates cannot be obtained, this method can be used to derive at least first-guess BSS masses.

A few BSSs in our sample have a median mass that exceeds $2M_{\mathrm{TO}}$, and could therefore be the product of more than two progenitors. Our uncertainties, however, are too large to conclusively confirm this results.

\acknowledgments

We thank the anonymous referee for useful comments that contributed to improve the presentation of the paper.

Based on observations with the NASA/ESA \textit{Hubble Space Telescope}, obtained at the Space Telescope Science Institute, which is operated by AURA, Inc., under NASA contract NAS 5-26555. 
This paper is part of the project COSMIC-LAB (``Globular Clusters as Cosmic Laboratories'') at the Physics and Astronomy Department of the Bologna University.

\vspace{5mm}
\facilities{\textit{HST}(ACS/HRC)}
\software{\texttt{emcee} (\citealt{foremanmackey13}); 
\texttt{Matplotlib} (\citealt{matplotlibref}); 
\texttt{NumPy} (\citealt{numpyref}); 
\texttt{pysynphot} (\citealt{pysynphotref}); 
\texttt{scipy} (\citealt{scipyref})}.

\end{document}